# The Capacity of Channels with Feedback

Sekhar Tatikonda
Yale University
New Haven, CT 06520
sekhar.tatikonda@yale.edu
203-432-4714

Sanjoy Mitter
Massachusetts Institute of Technology
Cambridge, MA, 02139
mitter@mit.edu

## Abstract

We introduce a general framework for treating channels with memory and feedback. First, we generalize Massey's concept of *directed information* [23] and use it to characterize the feedback capacity of general channels. Second, we present coding results for Markov channels. This requires determining appropriate sufficient statistics at the encoder and decoder. Third, a dynamic programming framework for computing the capacity of Markov channels is presented. Fourth, it is shown that the average cost optimality equation (ACOE) can be viewed as an implicit single-letter characterization of the capacity. Fifth, scenarios with simple sufficient statistics are described.

Keywords: Channel capacity, Feedback channels, Interconnection, Directed information, Markov channels, Stochastic control, Dynamic programming, Sufficient statistics

This research was supported by the National Science Foundation under award number 0430922: Capacity and Coding Techniques for Channels with Memory and Feedback.

# Contents

*A Sound Channel*
anagram of *Claude Shannon*

# 1 Introduction

This paper presents a general framework for proving coding theorems for channels with memory and feedback. The problem of optimal channel coding goes back to Shannon's original work [26]. The channel coding problem with feedback goes back to early work by Shannon, Dobrushin, Wolfowitz, and others [27], [11], [37]. Because of increased demand for wireless communication and networked systems there is a renewed interest in this problem. Feedback can increase the capacity of a noisy channel, decrease the complexity of the encoder and decoder, and reduce latency.

Recently Verdú and Han presented a very general formulation of the channel coding problem without feedback [33]. Specifically they provided a coding theorem for finite alphabet channels with arbitrary memory. They worked directly with the information density and provided a Feinstein-like lemma for the converse result. Here we generalize that formulation to the case of channels with feedback. In this case we require the use of code-functions as opposed to codewords. A code-function maps a message and the channel feedback information into a channel input symbol. Shannon introduced the use of code-functions, which he called strategies, in his work on transmitter side information [28]. Code-functions are also sometimes called codetrees [20].

We convert the channel coding problem with feedback into a new channel coding problem without feedback. The channel inputs in this new channel are code-functions. Unfortunately the space of code-functions can be quite complicated to work with. We show that we can work directly with the original space of channel inputs by making explicit the relationship between code-function distributions and channel input distributions. This relationship allows us to convert a mutual information optimization problem over code-function distributions into a directed information optimization problem over channel input distributions.

The concept of directed information was introduced by Massey who attributes it to Marko [23], [22]. Our work, in part, builds on the work of Kramer [19], [20]. He used the concept of directed information to prove capacity theorems for general discrete memoryless networks. These networks include the memoryless two-way channel and the memoryless multiple access channel. Here we examine single-user channels with memory and feedback.

There is a long history of work regarding Markov channels and feedback. Here we describe a few connections to that literature. Mushkin and Bar-David [24] examined the Gilbert-Elliot channel. Goldsmith and Variaya [15] examine non-ISI Markov channels without feedback. For the case of IID inputs and symmetric channels they introduce sufficient statistics that lead to a single-letter formula. We identify the appropriate sufficient statistics when feedback is available. In some sense the Markov channel with feedback problem is easier than the Markov channel without feedback problem because in the feedback case the decoder's information pattern is nested in the encoder's information pattern [35]. In this paper we do not treat noisy feedback. Viswanathan [34], Caire and Shamai [6], and Das and Narayan [9] all examine different classes of channels with memory and side-information at the transmitter and receiver. We present a general framework for treating Markov channels with ISI and feedback.

Many authors consider conditions that insure the Markov channel is *information stable* [25]. For example Cover and Pombra [7] show that Gaussian channels with feedback are always information stable. In addition some authors consider conditions that insure the Markov channel is *indecomposable* [14], [5]. In our work it is shown that solutions to the associated average cost optimality equation (ACOE) imply information stability. In addition the sufficient condition provided here for the existence of a solution to the ACOE implies a strong mixing property of the underlying Markov channel in the same way that indecomposability does.

We consider Markov channels with finite state, channel input, and channel output alphabets. But with the introduction of appropriate sufficient statistics we quickly find ourselves working with Markov channels over general alphabets and states. As shown by Csiszár [8], for example, treating general alphabets involve many technical issues that do not arise in the finite alphabet case.

Tatikonda first introduced the dynamic programming approach to computing directed information in his PhD thesis [29]. Yang, Kavcic, and Tatikonda have examined the case of finite state machine Markov channels in [38], [39]. Here we present a general framework that treats many Markov channels including finite state machine Markov channels.

In summary, the main contributions of this paper are:

(1) We generalize Massey's concept of *directed information* [23] and use it to characterize the feedback capacity of general channels.

(2) We present coding results for Markov channels. This requires determining appropriate sufficient statistics at the encoder and decoder.

(3) A dynamic programming framework for computing the capacity of Markov channels is presented.

(4) It is shown that the average cost optimality equation (ACOE) can be viewed as an implicit single-letter characterization of the capacity.

(5) Scenarios with simple sufficient statistics are described.

Preliminary versions of this work have appeared in [29], [30], [31], [32].

## 2  Feedback and Causality

Here we discuss some of the subtleties of feedback and causality inherent in the feedback capacity problem. The channel at time $t$ is modelled as a stochastic kernel $p(db_t \mid a^t, b^{t-1})$. Where $a^t = (a_1, ..., a_t)$ and $a^0 = \emptyset$. See figure one. The channel output is fed back to the encoder with delay one. At time $t$ the encoder takes the message and the past channel output symbols $B_1, ..., B_{t-1}$ and produces a channel input symbol $A_t$. At time $T$ the decoder takes all the channel output symbols, $B_1, ..., B_T$, and produces the decoded message. Hence

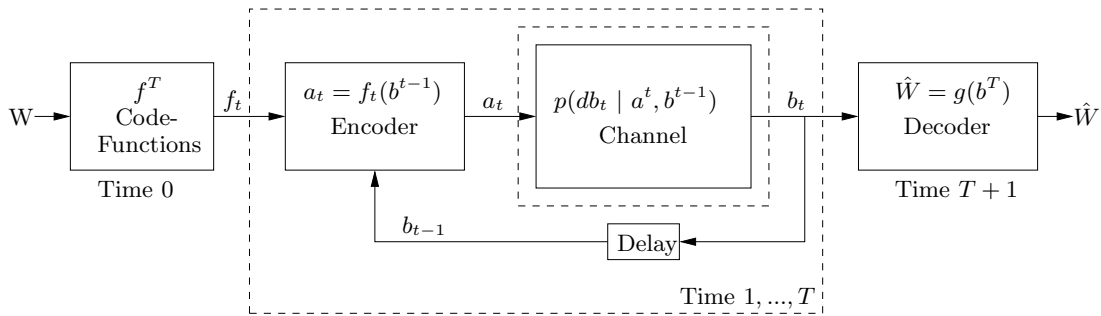

Figure 1: Interconnection

the time ordering of the variables is

$$\text{Message, } A_1, B_1, A_2, B_2, ..., A_T, B_T, \text{ Decoded Message.} \tag{1}$$

When there is no feedback, under suitable conditions, $\sup_{P(da^T)} I(A^T; B^T)$ characterizes the maximum number of messages one can send with small probability of decoding error. Our goal in this paper is to generalize this to the case of feedback. To that end we now mention some subtleties that will guide our approach.

*One should not supremize the mutual information $I(A^T, B^T)$ over the stochastic kernel* $p(da^T \mid b^T)$. We can factor $p(da^T \mid b^T) = \otimes_{t=1}^{T} p(da_t \mid a^{t-1}, b^T)$. This states that at time $t$ the channel input symbol $A_t$ is allowed to depend on the future channel output symbols $B_t^T$. This violates the causality implicit in our encoder description. In fact $p(da^T \mid b^T)$ is the posterior probability used by the decoder to decode the message at time $T$. Instead, as we will show, one should supremize the mutual information over the *directed stochastic kernel*: $\vec{p}(da^T \mid b^T) = \otimes_{t=1}^{T} p(da_t \mid a^{t-1}, b^{t-1})$. See definition 4.1.

*One should not use the stochastic kernel $p(db^T \mid a^T)$ as a model of the channel when there is feedback.* To compute the mutual information we need to work with the joint measure $P(da^T, db^T)$. In general it is not possible to find a joint measure consistent with the stochastic kernels: $\vec{p}(da^T \mid b^T)$ and $p(db^T \mid a^T)$. Instead, as we will show, the appropriate model for the channel when there is feedback is a sequence of stochastic kernels: $\{p(db_t \mid a^t, b^{t-1})\}_{t=1}^{T}$. See section 3.

*One should not use the mutual information $I(A^T; B^T)$ when there is feedback.* When there is feedback the conditional probabilities $P(db_t \mid a^T, b^{t-1}) \neq P(db_t \mid a^t, b^{t-1})$ almost surely under $P(da^T, db^T)$. Even though $A_{t+1}$ occurs after $B_t$ it still has a probabilistic influence on it. This is because under feedback $A_{t+1}$ is influenced by the past channel output $B_t$. To quote Massey [23], "statistical dependence, unlike causality, has no inherent directivity." The mutual information $I(A^T; B^T) = \sum_{t=1}^{T} I(A^T; B_t \mid B^{t-1})$. The information transmitted to the receiver at time $t$, given by $I(A^T; B_t \mid B^{t-1})$, depends on the future $A_{t+1}^T$. Instead, as we will show, we should use the *directed information*: $I(A^T \to B^T)$. See definition 4.2.

## 3 Channels with Feedback

In this section we formulate the feedback channel coding problem. We first introduce some notation. Let $p(dy \mid x)$ represent a stochastic kernel from the measurable spaces $\mathcal{X}$ to $\mathcal{Y}$. We say a stochastic kernel is *continuous* if for all continuous bounded functions $v$ on $\mathcal{Y}$ the function $\int v(y) p(dy \mid x)$ is a continuous and bounded function on $\mathcal{X}$. See appendix for definitions and properties of stochastic kernels.

Given a joint measure $P(dx, dy)$ we will use $P(Y = y \mid X = x)$ (or just $P(y \mid x)$) to represent the conditional probability (when it exists.) In general, lower case letters $p, q, r, ...$ will be used for stochastic kernels and upper case letter $P, Q, R, ...$ will be used for joint measures or conditional probabilities. Let $\mathcal{P}(\mathcal{X})$ represent the space of all probability measures on $\mathcal{X}$ endowed with the topology of weak convergence.

Capital letters, $A, B, X, Y, Z, ...$, will represent random variables and lower case letters, $a, b, x, y, z, ...$, will represent particular realizations. For the stochastic kernel $p(dy \mid x)$ we have $p(y \mid x)$ is a number. Given a joint measure $P_{X,Y}(dx, dy) = P_X(dx) \otimes p(dy \mid x)$ we have $p(y \mid X)$ is a random variable taking value $p(y \mid x)$ with probability $P_X(x)$, $p(Y \mid X)$ is a random variable taking value $p(y \mid x)$ with probability $P_{X,Y}(x, y)$, and $p(dy \mid X)$ is a random measure-valued element taking value $p(dy \mid x)$ with probability $P_X(x)$. Finally let the notation $X - Y - Z$ denote that the random elements $X, Y, Z$ form a Markov chain.

We are now ready to formulate the feedback channel coding problem. Let $\{A_t\}_{t=1}^T$ be random elements in the finite[1] set $\mathcal{A}$ with the power-set $\sigma$-algebra. These represent the channel inputs. Similarly let $\{B_t\}_{t=1}^T$ be random elements in the finite set $\mathcal{B}$ with the power-set $\sigma$-algebra. These represent the channel outputs. Let $\mathcal{A}^T$ and $\mathcal{B}^T$ represent the $T$-fold product spaces with appropriate product $\sigma$-algebras (where $T$ may be infinity.) We use "log" to represent logarithm base 2.

A *channel* is a family of stochastic kernels $\{p(db_t \mid a^t, b^{t-1})\}_{t=1}^T$. These channels are nonanticipative with respect to time-ordering (1) because the conditioning includes only $a^t, b^{t-1}$.

Let $\mathcal{F}_t$ be the set of all measurable maps $f_t : \mathcal{B}^{t-1} \to \mathcal{A}$ taking $b^{t-1} \mapsto a_t$. Endow $\mathcal{F}_t$ with the power-set $\sigma$-algebra. Let $\mathcal{F}^T = \prod_{t=1}^T \mathcal{F}_t$ denote the Cartesian product endowed with the product $\sigma$-algebra. Note that since $\mathcal{A}$ and $\mathcal{B}$ are finite the space $\mathcal{F}^T$ is at most countable. A *channel code-function* is an element $f^T = (f_1, ..., f_T) \in \mathcal{F}^T$. A distribution on $\mathcal{F}^T$ is given by a specification of a sequence of *code-function stochastic kernels* $\{p(df_t \mid f^{t-1})\}_{t=1}^T$. Specifically $P_{F^T}(df^T) = \otimes_{t=1}^T p(df_t \mid f^{t-1})$. We will use the notation $f^t(b^{t-1}) = (f_1, f_2(b_1), ..., f_t(b^{t-1}))$.

A *message set* is a set $\mathcal{W} = \{1, ..., M\}$. Let the distribution $P_W$ on the message set $\mathcal{W}$ be the uniform distribution. A *channel code*, is a set of $M$ channel code-functions denoted by $f^T[w]$, $w \in \mathcal{W}$. For message $w$ at time $t$ with channel feedback $b^{t-1}$ the channel encoder outputs $f_t[w](b^{t-1})$. A *channel code without feedback*, is a set of $M$ channel codewords denoted by $a^T[w]$, $w \in \mathcal{W}$. For message $w$ at time $t$ the channel encoder outputs $a_t[w]$ independent of the past channel outputs $b^{t-1}$.

---
[1] The methods in this paper can be generalized to channels with abstract alphabets.

A *channel decoder* is a map $g : \mathcal{B}^T \to \mathcal{W}$ taking $b^T \mapsto w$. The decoder waits till it observes all the channel outputs before reconstructing the input message. The order of events is shown in figure one.

**Definition 3.1** *A $(T, M, \epsilon)$ channel code over time horizon $T$ consists of $M$ code-functions, a channel decoder $g$, and an error probability satisfying: $\frac{1}{M}\sum_{w=1}^{M} \Pr(w \neq g(b^T) \mid w) \leq \epsilon$. A $(T, M, \epsilon)$ channel code without feedback is defined similarly with the restriction that we use $M$ codewords.*

In what follows the superscript "o" and "nfb" represent the words "operational" and "no feedback." Following [33] we define:

**Definition 3.2** *$R$ is an $\epsilon$-achievable rate if, for every $\delta > 0$ there exists, for all sufficiently large $T$, $(T, M, \epsilon)$ channel codes with rate $\frac{\log M}{T} > R - \delta$. The maximum $\epsilon - achievable$ rate is the called the $\epsilon-$capacity and denoted $C_\epsilon^o$. The operational channel capacity is defined as the maximal rate that is $\epsilon$-achievable for all $0 < \epsilon < 1$ and is denoted $C^o$. Analogous definitions for $C_\epsilon^{o,nfb}$ and $C^{o,nfb}$ hold for the case without feedback.*

Before continuing we quickly remark on some other formulations in the literature. Some authors work with different sets of channels for each blocklength $T$. See for example Dobrushin [12] and Verdú and Han [33]. In our context this would correspond to a different sequence of channels for each $T$: $\{p_T(db_t \mid a^t, b^{t-1})\}_{t=1}^T$. Theorem 5.1 below will continue to hold if we use channels of this form. But we do not need this level of generality to proceed with the Markov channels described in section 6.

Note that in definition 3.2 we are seeking a single number $C^o$ that is an achievable capacity for all sufficiently large $T$. Some authors instead, see [7] for example, seek a sequence of numbers $\{C_T^o\}$ such that there exists a sequence of channel codes $\{(T, 2^{TC_T^o}, \epsilon_T)\}$ with $\epsilon_T \to 0$. It will turn out that for the time-invariant Markov channels described in section 6 the notion of capacity described in definition 3.2 is the appropriate one. We will further elaborate on this point in section 4.1 after we have reviewed the concept of information stability.

### 3.1 Interconnection of Code-Functions to the Channel

Now we are ready to interconnect the pieces: channel, code-functions, encoder, and decoder. We follow Dobrushin's program and define a joint measure over the variables of interest that is consistent with the different components [12]. We will define a new channel without feedback that connects the code-functions to the channel outputs. Corollary 3.1 below shows that we can connect the messages directly to the channel output symbols.

Let $\{p(df_t \mid f^{t-1})\}_{t=1}^T$ be a sequence of code-function stochastic kernels with joint measure is $P_{F^T}(df^T) = \otimes_{t=1}^T p(df_t \mid f^{t-1})$ on $\mathcal{F}^T$. For example $P_{F^T}$ may be a distribution that places mass $1/M$ on each of $M$ different code-functions. Given a sequence of code-function stochastic kernels $\{p(df_t \mid f^{t-1})\}_{t=1}^T$ and a channel $\{p(db_t \mid a^t, b^{t-1})\}_{t=1}^T$ we want to construct a new channel that interconnects the random variables $F^T$ to the random variables $B^T$. We use "Q" to denote the new joint measure $Q(df^T, da^T, db^T)$ that we will construct.

The following three reasonable properties should hold for our new channel.

**Definition 3.3** *A measure $Q(df^T, da^T, db^T)$ is said to be* consistent *with the code-function stochastic kernels $\{p(df_t \mid f^{t-1})\}_{t=1}^T$ and the channel $\{p(db_t \mid a^t, b^{t-1})\}_{t=1}^T$ if for each t:*

(1) *There is no feedback to the code-functions in the new channel: The measure on $\mathcal{F}^T$ is chosen at time 0. Thus it cannot causally depend on the $A_t$'s and $B_t$'s. Thus for each t and all $f_t$, we have*

$$Q(f_t \mid F^{t-1} = f^{t-1}, \ A^{t-1} = a^{t-1}, B^{t-1} = b^{t-1}) = p(f_t \mid f^{t-1})$$

*for $Q$ almost all $f^{t-1}, a^{t-1}, b^{t-1}$.*

(2) *The channel input is a function of the past outputs: For each t, $A_t = F_t(B^{t-1})$ $Q-a.s.$ Said another way, for each t and all $a_t$, we have*

$$Q(a_t \mid F^t = f^t, \ A^{t-1} = a^{t-1}, B^{t-1} = b^{t-1}) = \delta_{\{f_t(b^{t-1})\}}(a_t)$$

*for $Q$ almost all $f^t, a^{t-1}, b^{t-1}$. Here $\delta$ is the Dirac measure.*

(3) *The new channel preserves the properties of the underlying channel: For each t, and all $b_t$, we have*

$$Q(b_t \mid F^t = f^t, \ A^t = a^t, \ B^{t-1} = b^{t-1}) = p(b_t \mid a^t, b^{t-1})$$

*for $Q$ almost all $f^t, a^t, b^{t-1}$.*

The following lemma shows that there exists a unique *consistent* measure $Q$ and provides the channel from $\mathcal{F}^T$ to $\mathcal{B}^T$.

**Lemma 3.1** *Given a sequence of code-function stochastic kernels $\{p(df_t \mid f^{t-1})\}_{t=1}^T$ and a channel $\{p(db_t \mid a^t, b^{t-1})\}_{t=1}^T$ there exists a unique consistent measure $Q(df^T, da^T, db^T)$ on $\mathcal{F}^T \times \mathcal{A}^T \times \mathcal{B}^T$. Furthermore the channel from $\mathcal{F}^T$ to $\mathcal{B}^T$ for each t and all $b_t$ is given by*

$$Q(b_t \mid F^t = f^t, \ B^{t-1} = b^{t-1}) = p(b_t \mid f^t(b^{t-1}), \ b^{t-1}) \tag{2}$$

*for $Q$ almost all $f^t, b^{t-1}$.*

**Proof:** Let $Q(df^T, da^T, db^T) = \bigotimes_{t=1}^T p(df_t \mid f^{t-1}) \otimes \delta_{\{f_t(b^{t-1})\}}(da_t) \otimes p(db_t \mid a^t, b^{t-1})$. For finite $T$ this measure exists (see Appendix A.1). An application of the Ionescu-Tulcea theorem shows that this measure exists for the $T = \infty$ case. Clearly this $Q$ is consistent and by construction it is unique.

Note that for each $(f^t, b^t)$ the joint measure can be decomposed as

$$\begin{aligned}
Q(f^t, b^t) &= \sum_{a^t} Q(f^t, a^t, b^t) \\
&= \sum_{a^t} \prod_{i=1}^t p(f_i \mid f^{i-1}) \ \delta_{\{f_i(b^{i-1})\}}(a_i) \ p(b_i \mid a^i, b^{i-1}) \\
&= p(b_t \mid f^t(b^{t-1}), b^{t-1}) \sum_{a^{t-1}} p(f_t \mid f^{t-1}) \prod_{i=1}^{t-1} p(f_i \mid f^{i-1}) \ \delta_{\{f_i(b^{i-1})\}}(a_i) \ p(b_i \mid a^i, b^{i-1}) \\
&= p(b_t \mid f^t(b^{t-1}), b^{t-1}) \ Q(f^t, b^{t-1})
\end{aligned}$$

Thus we have shown equation (2). $\square$

Hence for any sequence of code-function stochastic kernels $\{p(df_t \mid f^{t-1})\}_{t=1}^T$ the stochastic kernel $p(db_t \mid f^t(b^{t-1}), b^{t-1})$ can be chosen as a version of the regular conditional distribution $Q(db_t \mid F^t = f^t, B^{t-1} = b^{t-1})$. Thus the stochastic kernels $\{p(db_t \mid f^t(b^{t-1}), b^{t-1})\}_{t=1}^T$ can be viewed as the channel from $\mathcal{F}^T$ to $\mathcal{B}^T$. Note that the dependence is on $f^t(b^{t-1})$ and not $f^t$. We will see in section 5 that this observation will greatly simplify computation.

The almost sure qualifier in equation (2) comes from the fact that $Q(f^t, b^{t-1})$ may equal zero for some $f^t, b^{t-1}$. This can happen, for example, if either $f^t$ has zero probability of appearing under $P_{F^T}(df^T)$ or $b^{t-1}$ has zero probability of appearing under the channel $\{p(db_t \mid a^t, b^{t-1})\}$.

A distribution $P_W$ on $\mathcal{W}$ induces a measure $P_{F^T}$ on $\mathcal{F}^T$. Hence:

**Corollary 3.1** *A distribution $P_W$ on $\mathcal{W}$, a channel code $\{f^T[w]\}_{w=1}^M$, and the channel $\{p(db_t \mid a^t, b^{t-1})\}_{t=1}^T$ uniquely define a measure $Q(dw, da^T, db^T)$ on $\mathcal{W} \times \mathcal{A}^T \times \mathcal{B}^T$. Furthermore the channel from $\mathcal{W}$ to $\mathcal{B}^T$ for each $t$ and all $b_t$ is given by*

$$Q(b_t \mid W = w, \ B^{t-1} = b^{t-1}) = p(b_t \mid f^t[w](b^{t-1}), \ b^{t-1})$$

*for $Q$ almost all $w, b^{t-1}$.*

## 4 Directed Information

As discussed in section 2 the traditional mutual information is insufficient for dealing with channels with feedback. Here we generalize Massey's notion of *directed information* to take into account any time-ordering of the random variables of interest.

**Definition 4.1** *We are given a sequence of stochastic kernels $\{p(da_i \mid a^{i-1})\}_{i=1}^N$. Let $I = \{i_1, ..., i_K\} \subseteq \{1, ..., N\}$ where $1 \leq i_1 < i_2 < ... < i_K \leq N$. Let $I^c = \{1, ..., N\} \setminus I$. Let $A^I = (A_{i_1}, ..., A_{i_K})$. Define $A^{I^c}$ similarly. Then the* directed stochastic kernel *of $A^I$ with respect to $A^{I^c}$ is*

$$\vec{p}_{A^I \mid A^{I^c}}(da^I \mid a^{I^c}) = \bigotimes_{k=1}^K p_{A_{i_k} \mid A^{i_k-1}}(da_{i_k} \mid a^{i_k-1}).$$

For each $a^{I^c}$ the directed stochastic kernel $\vec{p}_{A^I \mid A^{I^c}}(dA^I \mid a^{I^c})$ is a well defined measure. For example:

$$\int f(a_1, ..., a_N) \vec{p}_{A^I \mid A^{I^c}}(da^I \mid a^{I^c})$$
$$= \int p(da_{i_1} \mid a^{i_1-1}) \int p(da_{i_2} \mid a^{i_2-1}) \cdots \int p(da_{i_k} \mid a^{i_k-1}) f(a_1, ..., a_N)$$

for all bounded functions $f$ measurable with respect to the product $\sigma$-algebra on $\mathcal{A}^N$. Note that this integral is a measurable function of $a^{I^c}$.

One needs to be careful when computing the marginals of a directed stochastic kernel. For example, if we are given $p(da_1)$, $p(da_2 \mid a_1)$, and $p(da_3 \mid a_1, a_2)$ with the resulting joint measure $P(da_1, da_2, da_3)$ then with the obvious time ordering:

$$\sum_{a_1 \in \mathcal{A}} \vec{p}(a_1, a_3 \mid a_2) = \sum_{a_1 \in \mathcal{A}} (p(a_3 \mid a_1, a_2) p(a_1)) \neq P(a_3 \mid a_2) \quad P-\text{almost all } a_1, a_2, a_3$$

unless $A_1 - A_2 - A_3$ forms a Markov chain under $P$. Here $P(a_3 \mid a_2)$ corresponds to the conditional probability under $P$.

**Definition 4.2** *Given a sequence of stochastic kernels $\{p(da_i \mid a^{i-1})\}_{i=1}^N$ and $I \subseteq \{1,...,N\}$, the* directed information *is defined as*

$$I(A^I \to A^{I^c}) = D(P_{A^I, A^{I^c}} \mid \vec{P}_{A^I \mid A^{I^c}} P_{A^{I^c}}) \tag{3}$$

where $D(\cdot \mid \cdot)$ is the divergence, $P_{A^I, A^{I^c}}(da^I, da^{I^c}) = \vec{p}_{A^{I^c} \mid A^I}(da^{I^c} \mid a^I) \otimes \vec{p}_{A^I \mid A^{I^c}}(da^I \mid a^{I^c})$, and $\vec{P}_{A^I \mid A^{I^c}} P_{A^{I^c}}(da^I, da^{I^c}) = \vec{p}_{A^I \mid A^{I^c}}(da^I \mid a^{I^c}) \otimes P_{A^{I^c}}(da^{I^c})$ (here $P_{A^{I^c}}(da^{I^c})$ is the marginal of $P_{A^I, A^{I^c}}(da^I, da^{I^c})$.)

We can recover Massey's definition of directed information [23] by applying definition 4.2 to $A^I = A^T$ and $A^{I^c} = B^T$ with the time-ordering given in (1): $I(A^T \to B^T) = \sum_{t=1}^T I(A^t; B_t \mid B^{t-1})$. Unlike the chain rule for mutual information the superscript on $A$ in the summation is "$t$" and not "$T$". From definition 4.2 one can easily show:

$$I(A^T \to B^T) = E\left[\log \frac{\vec{p}_{B^T \mid A^T}(B^T \mid A^T)}{P_{B^T}(B^T)}\right] = E\left[\log \frac{p_{A^T \mid B^T}(A^T \mid B^T)}{\vec{p}_{A^T \mid B^T}(A^T \mid B^T)}\right]$$

where the stochastic kernel $p_{A^T \mid B^T}(da^T \mid b^T)$ is a version of the conditional distribution $P(da^T \mid b^T)$. The second equality shows that the directed information is the ratio between the posterior distribution and a "causal" prior distribution.

Note that $I(A^T; B^T) = E\left[\log \frac{\vec{p}(B^T \mid A^T) \vec{p}(A^T \mid B^T)}{P(B^T) P(A^T)}\right] = I(A^T \to B^T) + I(B^T \to A^T)$. By definition 4.2 and time-ordering (1) we have $I(B^T \to A^T) = \sum_{t=1}^T I(A_t; B^{t-1} \mid A^{t-1})$. There is no feedback if and only if $A_t - A^{t-1} - B^{t-1}$ forms a Markov chain under $P$. Hence $I(B^T \to A^T) = 0$. There is no "information" flowing from the receiver to the transmitter. Because divergence is nonnegative we can conclude that $I(A^T; B^T) \geq I(A^T \to B^T)$ with equality if and only if there is no feedback [23], [19].

## 4.1 Information Density, Directed Information, and Capacity

When computing the capacity of a channel it will turn out that we will need to know the convergence properties of the random variables $\frac{1}{T} \log \frac{P_{A^T, B^T}(A^T, B^T)}{\vec{P}_{A^T \mid B^T} P_{B^T}(A^T, B^T)}$. This is the *normalized information density* discussed in [33] suitably generalized to treat feedback. If there are reasonable regularity properties, like information stability (see below), then these random variables will converge in probability to a deterministic limit. In the absence of any such structure we are forced to follow Verdú and Han's lead and define the following "floor" and "ceiling" limits [33].

The *limsup in probability* of a sequence of random variables $\{X_t\}$ is defined as the smallest extended real number $\alpha$ such that $\forall \epsilon > 0 \lim_{t \to \infty} \Pr[X_t \geq \alpha + \epsilon] = 0$. The *liminf in probability* of a sequence of random variables $\{X_t\}$ is defined as the largest extended real number $\alpha$ such that $\forall \epsilon > 0 \lim_{t \to \infty} \Pr[X_t \leq \alpha - \epsilon] = 0$.

Let $\vec{i}(a^T; b^T) = \log \frac{P_{A^T, B^T}(a^T, b^T)}{\vec{P}_{A^T|B^T} P_{B^T}(a^T, b^T)}$. For a sequence of joint measures $\{P_{A^T, B^T}\}_{T=1}^{\infty}$ let

$$\underline{I}(A \to B) = \liminf_{\text{in prob}} \frac{1}{T} \vec{i}(A^T; B^T) \quad \text{and} \quad \overline{I}(A \to B) = \limsup_{\text{in prob}} \frac{1}{T} \vec{i}(A^T; B^T).$$

**Lemma 4.1** *For any sequence of joint measures $\{P_{A^T, B^T}\}_{T=1}^{\infty}$ we have*

$$\underline{I}(A \to B) \leq \liminf_{T \to \infty} \frac{1}{T} I(A^T \to B^T) \leq \limsup_{T \to \infty} \frac{1}{T} I(A^T \to B^T) \leq \overline{I}(A \to B)$$

**Proof:** See the appendix A.2. □

We now extend Pinsker's [25] notion of information stability. A given sequence of joint measures $\{P_{A^T, B^T}\}_{T=1}^{\infty}$ is *directed information stable* if $\lim_{T \to \infty} P\left(\left|\frac{\vec{i}(A^T; B^T)}{I(A^T \to B^T)} - 1\right| > \epsilon\right) = 0 \quad \forall \epsilon > 0$. The following lemma shows that directed information stability implies $\frac{1}{T}\vec{i}(a^T; b^T)$ concentrates around its mean $\frac{1}{T} I(A^T \to B^T)$. Note that this mean need not necessarily converge.

**Lemma 4.2** *If the sequence of joint measures $\{P_{A^T, B^T}\}_{T=1}^{\infty}$ is directed information stable then $\underline{I}(A \to B) = \liminf_{T \to \infty} \frac{1}{T} I(A^T \to B^T) \leq \limsup_{T \to \infty} \frac{1}{T} I(A^T \to B^T) = \overline{I}(A \to B)$.*

**Proof:** Directed information stability implies

$$\lim_{T \to \infty} P\left(\left|\frac{1}{T}\vec{i}(A^T; B^T) - \frac{1}{T} I(A^T \to B^T)\right| > \frac{1}{T} I(A^T \to B^T) \epsilon \right) = 0 \quad \forall \epsilon > 0.$$

Because $\mathcal{B}$ is finite we know $\frac{1}{T} I(A^T \to B^T) \leq \log |\mathcal{B}|$ hence

$$\lim_{T \to \infty} P\left(\left|\frac{1}{T}\vec{i}(A^T; B^T) - \frac{1}{T} I(A^T \to B^T)\right| > \epsilon \right) = 0 \quad \forall \epsilon > 0.$$

This observation along with lemma 4.1 proves the lemma. □

To compute the different "information" measures we need to determine the joint measure $P_{A^T, B^T}(da^T, db^T)$. This can be done if we are given a channel $\{p(db_t \mid a^t, b^{t-1})\}_{t=1}^{T}$ and we specify a sequence of kernels $\{p(da_t \mid a^{t-1}, b^{t-1})\}_{t=1}^{T}$.

**Definition 4.3** *A* channel input distribution *is a sequence of kernels $\{p(da_t \mid a^{t-1}, b^{t-1})\}_{t=1}^{T}$. A* channel input distribution without feedback *is a channel input distribution with the further condition that for each $t$ the kernel $p(da_t \mid a^{t-1}, b^{t-1})$ is independent of $b^{t-1}$. (Specifically $p(da_t \mid a^{t-1}, b^{t-1}) = p(da_t \mid a^{t-1}, \tilde{b}^{t-1}) \quad \forall b^{t-1}, \tilde{b}^{t-1}$.)*

Let $\mathcal{D}_T = \{\{p(da_t \mid a^{t-1}, b^{t-1})\}_{t=1}^{T}\}$ be the set of all channel input distributions. Let $\mathcal{D}_T^{\text{nfb}} \subset \mathcal{D}_T$ be the set of channel input distributions without feedback. We now define the directed information optimization problems. Fix a channel $\{p(db_t \mid a^t, b^{t-1})\}$. For finite $T$ let

$$C_T = \sup_{\mathcal{D}_T} \frac{1}{T} I(A^T \to B^T) \quad \text{and} \quad C_T^{\text{nfb}} = \sup_{\mathcal{D}_T^{\text{nfb}}} \frac{1}{T} I(A^T \to B^T) = \sup_{\mathcal{D}_T^{\text{nfb}}} \frac{1}{T} I(A^T; B^T).$$

For the infinite horizon case let

$$C = \sup_{\{\mathcal{D}_T\}_{T=1}^\infty} \underline{I}(A \to B) \quad \text{and} \quad C^{\text{nfb}} = \sup_{\{\mathcal{D}_T^{\text{nfb}}\}_{T=1}^\infty} \underline{I}(A \to B) = \sup_{\{\mathcal{D}_T^{\text{nfb}}\}_{T=1}^\infty} \underline{I}(A; B) \quad (4)$$

Verdú and Han proved the following theorem for the case without feedback [33].

**Theorem 4.1** *For channels without feedback $C^{o,\text{nfb}} = C^{\text{nfb}}$.*

In a certain sense we already have the solution to the coding problem for channels with feedback. Specifically lemma 3.1 tells us that the feedback channel problem is equivalent to a new channel coding problem without feedback. This new channel is from $\mathcal{F}^T$ to $\mathcal{B}^T$ and has channel kernels defined by equation (2). Thus we can directly apply theorem 4.1 to this new channel.

This can be a very complicated problem to solve. We would have to optimize the mutual information over distributions on code functions. The directed information optimization problem can often be simpler. One reason is that we can work directly on the original $\mathcal{A}^T \times \mathcal{B}^T$ space and not on the $\mathcal{F}^T \times \mathcal{B}^T$ space. The second half of this paper describes a stochastic control approach to solving this optimization. In the next section, though, we present the feedback coding theorem.

## 5 Coding Theorem for Channels with Feedback

In this section we prove the following theorem:

**Theorem 5.1** *For channels with feedback $C^o = C$.*

We first give a high-level summary of the issues involved. The converse part is straightforward. For any channel code and channel we know by lemma 3.1 that there exists a unique consistent measure $Q(df^T, da^T, db^T)$. From this measure we can compute the induced channel input distribution $\{q(da_t \mid a^{t-1}, b^{t-1})\}_{t=1}^T$. (These stochastic kernels are a version of the appropriate conditional probabilities.) Now $\{q(da_t \mid a^{t-1}, b^{t-1})\}_{t=1}^T \in \mathcal{D}_T$ but it need not be the supremizing channel input distribution. Thus the directed information under the induced channel input distribution may be less than the directed information under the supremizing channel input distribution. This is how we will show $C^o \leq C$.

The direct part is the interesting part of the theorem 5.1. Here we take the optimizing channel input distribution $\{p(da_t \mid a^{t-1}, b^{t-1})\}_{t=1}^T$ and construct a sequence of code-function stochastic kernels $\{p(df_t \mid f^{t-1})\}_{t=1}^T$. We then prove the direct part of the coding theorem for the channel from $\mathcal{F}^T$ to $\mathcal{B}^T$ by the usual techniques for channels without feedback. By a suitable construction of $P_{F^T}$ it can be shown that the induced channel input distribution equals the original channel input distribution.

In section 5.1 we provide the necessary technical lemmas to characterize the relationship between code-function distributions and channel input distributions, in section 5.2 we prove theorem 5.1, and in section 5.3 we generalize the theorem to more general information patterns at the encoder.

## 5.1 Main Technical Lemmas

We first discuss the channel input distribution induced by a given code-function distribution. Define the $graph(f_t) = \{(b^{t-1}, a_t) : f_t(b^{t-1}) = a_t\} \subset \mathcal{B}^{t-1} \times \mathcal{A}$. Let $\Upsilon_t(b^{t-1}, a_t) = \{f_t : (b^{t-1}, a_t) \in \text{graph}(f_t)\}$ and $\Upsilon^t(b^{t-1}, a^t) = \{f^t : (b^{j-1}, a_j) \in \text{graph}(f_j), \ j = 1, ..., t\}$.

In lemma 3.1 we showed the channel from $\mathcal{F}^T$ to $\mathcal{B}^T$ depends only on the channel from $\mathcal{A}^T$ to $\mathcal{B}^T$. Hence for each $t$ and all $b_t$, we have

$$\begin{aligned} Q(b_t \mid F^t = f^t, \ B^{t-1} = b^{t-1}) &= p(b_t \mid f^t(b^{t-1}), \ b^{t-1}) \quad Q-\text{almost all } f^t, b^{t-1} \\ &= p(b_t \mid \tilde{f}^t(b^{t-1}), \ b^{t-1}) \quad \forall \tilde{f}^t \in \Upsilon^t(b^{t-1}, f^t(b^{t-1})) \end{aligned}$$

We now show that the induced channel input distribution only depends on the sequence of code-function stochastic kernels $\{p(df_t \mid f^{t-1})\}_{t=1}^T$.

**Lemma 5.1** *We are given a sequence of code-function stochastic kernels $\{p(df_t \mid f^{t-1})\}_{t=1}^T$, a channel $\{p(db_t \mid a^t, b^{t-1})\}_{t=1}^T$, and a consistent joint measure $Q(df^T, da^T, db^T)$. Then the induced channel input distribution is, for each $t$ and all $a_t$, given by*

$$Q(a_t \mid a^{t-1}, b^{t-1}) = P_{F^T}\left(\Upsilon_t(b^{t-1}, a_t) \mid \Upsilon^{t-1}(b^{t-2}, a^{t-1})\right) \tag{5}$$

*for $Q$ almost all $a^{t-1}, b^{t-1}$. Where $P_{F^T}(df^T) = \otimes_{t=1}^T p(df_t \mid f^{t-1})$.*

**Proof:** Note $P_{F^T}(\Upsilon^{t-1}(b^{t-2}, a^{t-1})) = Q(\Upsilon^{t-1}(b^{t-2}, a^{t-1})) \geq Q(\Upsilon^{t-1}(b^{t-2}, a^{t-1}), a^{t-1}, b^{t-1})$
$= Q(a^{t-1}, b^{t-1})$. Thus $Q(a^{t-1}, b^{t-1}) > 0$ implies $P_{F^T}(\Upsilon^{t-1}(b^{t-2}, a^{t-1})) > 0$. Hence the right hand side of equation (5) exists $Q$-almost surely.

We now prove the correctness of equation (5). For each $t$ and $(a^{t-1}, b^{t-1})$ such that $Q(a^{t-1}, b^{t-1}) > 0$ we have

$$Q(a^t, b^{t-1})$$

$$= \sum_{f^t} \left(\prod_{i=1}^t p(f_i \mid f^{i-1}) \delta_{\{f_i(b^{i-1})\}}(a_i)\right) \vec{p}(b^{t-1} \mid a^{t-1})$$

$$\stackrel{(a)}{=} \sum_{f^t \in \Upsilon^t(b^{t-1}, a^t)} \left(\prod_{i=1}^{t-1} p(f_i \mid f^{i-1})\right) \vec{p}(b^{t-1} \mid a^{t-1})$$

$$\stackrel{(b)}{=} P_{F^T}(\Upsilon_t(b^{t-1}, a_t) \mid \Upsilon^{t-1}(b^{t-2}, a^{t-1})) P_{F^T}(\Upsilon^{t-1}(b^{t-2}, a^{t-1})) \vec{p}(b^{t-1} \mid a^{t-1})$$

$$= P_{F^T}(\Upsilon_t(b^{t-1}, a_t) \mid \Upsilon^{t-1}(b^{t-2}, a^{t-1})) \sum_{f^{t-1}} \left(\prod_{i=1}^{t-1} p(f_i \mid f^{i-1}) \delta_{\{f_i(b^{i-1})\}}(a_i)\right) \vec{p}(b^{t-1} \mid a^{t-1})$$

$$= P_{F^T}(\Upsilon_t(b^{t-1}, a_t) \mid \Upsilon^{t-1}(b^{t-2}, a^{t-1})) Q(a^{t-1}, b^{t-1})$$

where (a) follows because $\vec{p}(b^{t-1} \mid a^{t-1})$ does not depend on $f^{t-1}$ and the delta functions $\{\delta_{f_i(b^{i-1})}(a_i)\}$ restrict the sum over $f^{t-1}$. Line (b) follows because $Q(a^{t-1}, b^{t-1}) > 0$ and hence the conditional probability exists. □

The almost sure qualifier in equation (5) comes from fact that $Q(a^{t-1}, b^{t-1})$ may equal zero for some $a^{t-1}, b^{t-1}$. This can happen, for example, if $P_{F^T}(df^T)$ puts zero mass on those $f^{t-1}$ that produce $a^{t-1}$ from $b^{t-1}$ or if $b^{t-1}$ has zero probability of appearing under the channel $\{p(db_t \mid a^t, b^{t-1})\}$.

We now show the equivalence of the directed information measures for both the "$\mathcal{F}^T - \mathcal{B}^T$" and the "$\mathcal{A}^T - \mathcal{B}^T$" channels.

**Lemma 5.2** *For each finite $T$ and every consistent joint measure $Q(df^T, da^T, db^T)$ we have*

$$\frac{Q_{F^T, B^T}(F^T, B^T)}{Q_{F^T} Q_{B^T}(F^T, B^T)} = \frac{Q_{A^T, B^T}(A^T, B^T)}{\vec{Q}_{A^T \mid B^T} Q_{B^T}(A^T, B^T)} \quad Q - a.s. \tag{6}$$

*hence $I(F^T; B^T) = I(A^T \to B^T)$. Furthermore, if given a sequence of consistent measures $\{Q(df^T, da^T, db^T)\}_{T=1}^{\infty}$, then $\underline{I}(F; B) = \underline{I}(A \to B)$.*

**Proof:** Fix $T$ finite. Then for every $(f^T, a^T, b^T)$ such that $Q(f^T, a^T, b^T) > 0$ we have

$$\begin{aligned}
\frac{Q_{F^T, B^T}(f^T, b^T)}{Q_{F^T} Q_{B^T}(f^T, b^T)} &= \frac{\sum_{\tilde{a}^T} Q(f^T, \tilde{a}^T, b^T)}{Q_{B^T}(b^T) Q_{F^T}(f^T)} \\
&= \frac{\sum_{\tilde{a}^T} \prod_{t=1}^{T} p(f_t \mid f^{t-1}) \delta_{\{f_t(b^{t-1})\}}(\tilde{a}_t) p(b_t \mid \tilde{a}^t, b^{t-1})}{Q_{B^T}(b^T) Q_{F^T}(f^T)} \\
&= \frac{P_{F^T}(f^T) \vec{p}_{B^T \mid A^T}(b^T \mid a^T)}{Q_{B^T}(b^T) Q_{F^T}(f^T)} \\
&\stackrel{(a)}{=} \frac{\vec{p}_{B^T \mid A^T}(b^T \mid a^T) P_{F^T}(\Upsilon(b^{T-1}, a^T))}{Q_{B^T}(b^T) \vec{q}_{A^T \mid B^T}(a^T \mid b^T)} \\
&= \frac{\vec{p}_{B^T \mid A^T}(b^T \mid a^T) \sum_{\tilde{f}^T} \prod_{t=1}^{T} p(\tilde{f}_t \mid \tilde{f}^{t-1}) \delta_{\{\tilde{f}_t(b^{t-1})\}}(a_t)}{Q_{B^T}(b^T) \vec{q}_{A^T \mid B^T}(a^T \mid b^T)} \\
&= \frac{\sum_{\tilde{f}^T} \prod_{t=1}^{T} p(\tilde{f}_t \mid \tilde{f}^{t-1}) \delta_{\{\tilde{f}_t(b^{t-1})\}}(a_t) p(b_t \mid a^t, b^{t-1})}{Q_{B^T}(b^T) \vec{q}_{A^T \mid B^T}(a^T \mid b^T)} \\
&= \frac{\sum_{\tilde{f}^T} Q(\tilde{f}^T, a^T, b^T)}{Q_{B^T}(b^T) \vec{q}_{A^T \mid B^T}(a^T \mid b^T)} \\
&= \frac{Q_{A^T, B^T}(a^T, b^T)}{\vec{Q}_{A^T \mid B^T} Q_{B^T}(a^T, b^T)}
\end{aligned}$$

where (a) follows because the $Q$ marginal $Q(df^T) = P_{F^T}(df^T)$ and for $Q(f^T, a^T, b^T) > 0$ lemma 5.1 shows $P_{F^T}(\Upsilon(b^{T-1}, a^T)) = \vec{q}_{A^T \mid B^T}(a^T \mid b^T)$.

Furthermore, if given a sequence of consistent measures $\{Q(df^T, da^T, db^T)\}_{T=1}^{\infty}$, equation (6) states that for each $T$ the random variables on the left hand side and right hand side are almost surely equal. Hence $\underline{I}(F; B) = \underline{I}(A \to B)$. □

We have shown how a code-function distribution induces a channel input distribution. As we discussed in the introduction to this section, we would like to choose a channel input distribution, $\{p(da_t \mid a^{t-1}, b^{t-1})\}_{t=1}^T$, and construct a sequence of code-function stochastic kernels, $\{p(df_t \mid f^{t-1})\}_{t=1}^T$, such that the resulting induced channel input distribution, $\{q(da_t \mid a^{t-1}, b^{t-1})\}_{t=1}^T$, equals the chosen channel input distribution. This is shown pictorially:

$$\boxed{\{p(da_t \mid a^{t-1}, b^{t-1})\}_{t=1}^T} \longrightarrow \boxed{\{p(df_t \mid f^{t-1})\}_{t=1}^T} \longrightarrow \boxed{\{q(da_t \mid a^{t-1}, b^{t-1})\}_{t=1}^T}$$

The first arrow represents the construction of the code-function distribution from the chosen channel input distribution. The second arrow is described by the result in lemma 5.1. Lemma 5.2 states that $\underline{I}_Q(F; B) = \underline{I}_Q(A \to B)$. Let $P$ correspond to the joint measure determined by the left channel input distribution in the diagram. If we can find conditions such that the induced channel input distribution equals the chosen channel input distribution then $\underline{I}_Q(A \to B) = \underline{I}_P(A \to B)$. Consequently $\underline{I}_Q(F; B) = \underline{I}_P(A \to B)$.

**Definition 5.1** *We call a sequence of code-function stochastic kernels $\{p(df_t \mid f^{t-1})\}_{t=1}^T$, with resulting joint measure $P_{F^T}(df^T)$, good with respect to the channel input distribution $\{p(da_t \mid a^{t-1}, b^{t-1})\}_{t=1}^T$ if for each $t$ and all $a^t, b^{t-1}$ we have*

$$P_{F^T}(\Upsilon^t(b^{t-1}, a^t)) = \vec{p}(a^t \mid b^{t-1}).$$

Lemma 5.4 below shows good code-function distributions exists. Before proving that we show the equivalence of the chosen and induced channel input distributions when a good code-function distribution is used.

**Lemma 5.3** *We are given a sequence of code-function stochastic kernels $\{p(df_t \mid f^{t-1})\}_{t=1}^T$, a channel $\{p(db_t \mid a^t, b^{t-1})\}_{t=1}^T$, and a consistent joint measure $Q(df^T, da^T, db^T)$. We are also given a channel input distribution $\{r(da_t \mid a^{t-1}, b^{t-1})\}_{t=1}^T$. The induced channel input distribution satisfies for each $t$ and all $a_t$*

$$Q(a_t \mid a^{t-1}, b^{t-1}) = r(a_t \mid a^{t-1}, b^{t-1}) \text{ for } Q \text{ almost all } a^{t-1}, b^{t-1} \qquad (7)$$

*if and only if the sequence of code-function stochastic kernels $\{p(df_t \mid f^{t-1})\}_{t=1}^T$ is good with respect to $\{r(da_t \mid a^{t-1}, b^{t-1})\}_{t=1}^T$.*

**Proof:** Note that for each $t$ and all $a_t$:

$$\begin{aligned}
Q(a_t \mid a^{t-1}, b^{t-1}) &\stackrel{(a)}{=} P_{F^T}(\Upsilon_t(b^{t-1}, a_t) \mid \Upsilon^{t-1}(b^{t-2}, a^{t-1})) \quad Q - \text{almost all } a^{t-1}, b^{t-1} \\
&\stackrel{(b)}{=} \frac{\vec{r}(a^t \mid b^{t-1})}{\vec{r}(a^{t-1} \mid b^{t-2})} \quad Q - \text{almost all } a^{t-1}, b^{t-1} \\
&= r(a_t \mid a^{t-1}, b^{t-1}) \quad Q - \text{almost all } a^{t-1}, b^{t-1}
\end{aligned}$$

where (a) follows from lemma 5.1 and (b) follows from definition 5.1. □

**Lemma 5.4** *For any channel input distribution $\{p(da_t \mid a^{t-1}, b^{t-1})\}_{t=1}^T$ there exists a sequence of code-functions stochastic kernels that are good with respect to it.*

**Proof:** For all $f^t$ define $p(f_t \mid f^{t-1})$ as follows
$$p(f_t \mid f^{t-1}) = \prod_{b^{t-1}} p(f_t(b^{t-1}) \mid f^{t-1}(b^{t-2}), b^{t-1}) \tag{8}$$

We first show that $p(f_t \mid f^{t-1})$ defined in equation (8) is a stochastic kernel. Note that for each $t$ and all $f^{t-1}$ we have

$$\sum_{f_t} p(f_t \mid f^{t-1})$$
$$= \sum_{f_t} \prod_{b^{t-1}} p(f_t(b^{t-1}) \mid f^{t-1}(b^{t-2}), b^{t-1})$$
$$= \sum_{a_t} \sum_{f_t \in \Upsilon_t(\bar{b}^{t-1}, a_t)} \prod_{b^{t-1}} p(f_t(b^{t-1}) \mid f^{t-1}(b^{t-2}), b^{t-1})$$
$$= \sum_{a_t} \sum_{f_t \in \Upsilon_t(\bar{b}^{t-1}, a_t)} p(f_t(\bar{b}^{t-1}) \mid f^{t-1}(\bar{b}^{t-2}), \bar{b}^{t-1}) \prod_{b^{t-1} \neq \bar{b}^{t-1}} p(f_t(b^{t-1}) \mid f^{t-1}(b^{t-2}), b^{t-1})$$
$$= \sum_{a_t} p(a_t \mid f^{t-1}(\bar{b}^{t-2}), \bar{b}^{t-1}) \sum_{f_t \in \Upsilon_t(\bar{b}^{t-1}, a_t)} \prod_{b^{t-1} \neq \bar{b}^{t-1}} p(f_t(b^{t-1}) \mid f^{t-1}(b^{t-2}), b^{t-1})$$
$$\stackrel{(a)}{=} \prod_{b^{t-1}} \sum_{a_t} p(a_t \mid f^{t-1}(b^{t-2}), b^{t-1})$$
$$= 1$$

where (a) follows by repeating the previous step for each $b^{t-1}$. In short, the sum is over *all* functions $f_t : \mathcal{B}^{t-1} \to \mathcal{A}$. Hence the sum over $f_t$ can be viewed as a sum over all assignments of $a_t$'s to each choice of $b^{t-1}$. Then the sum of products can be written as a product of sums.

We now show by induction that for each $t$ and all $a^t, b^{t-1}$ we have $P_{FT}(\Upsilon^t(b^{t-1}, a^t)) = \vec{p}(a^t \mid b^{t-1})$. For $t = 1$ we have $P_{FT}(\Upsilon_1(a_1)) = \sum_{f_1 \in \Upsilon_1(a_1)} p(f_1) = p(a_1)$. For $t + 1$ we have

$$P_{FT}(\Upsilon^{t+1}(b^t, a^{t+1}))$$
$$= \sum_{f^t \in \Upsilon^t(b^{t-1}, a^t)} \prod_{i=1}^{t} p(f_i \mid f^{i-1}) \sum_{f_{t+1} \in \Upsilon_{t+1}(b^t, a_{t+1})} p(f_{t+1} \mid f^t)$$
$$= \sum_{f^t \in \Upsilon^t(b^{t-1}, a^t)} \prod_{i=1}^{t} p(f_i \mid f^{i-1}) \sum_{f_{t+1} \in \Upsilon_{t+1}(b^t, a_{t+1})} \prod_{\tilde{b}^t} p(f_{t+1}(\tilde{b}^t) \mid f^t(\tilde{b}^{t-1}), \tilde{b}^t)$$
$$= \sum_{f^t \in \Upsilon^t(b^{t-1}, a^t)} \prod_{i=1}^{t} p(f_i \mid f^{i-1}) \sum_{f_{t+1} \in \Upsilon_{t+1}(b^t, a_{t+1})} p(f_{t+1}(b^t) \mid f^t(b^{t-1}), b^t) \prod_{\tilde{b}^t \neq b^t} p(f_{t+1}(\tilde{b}^t) \mid f^t(\tilde{b}^{t-1}), \tilde{b}^t)$$
$$\stackrel{(a)}{=} \sum_{f^t \in \Upsilon^t(b^{t-1}, a^t)} \prod_{i=1}^{t} p(f_i \mid f^{i-1}) p(a_{t+1} \mid a^t, b^t) \sum_{f_{t+1} \in \Upsilon_{t+1}(b^t, a_{t+1})} \prod_{\tilde{b}^t \neq b^t} p(f_{t+1}(\tilde{b}^t) \mid f^t(\tilde{b}^{t-1}), \tilde{b}^t)$$
$$\stackrel{(b)}{=} \sum_{f^t \in \Upsilon^t(b^{t-1}, a^t)} \prod_{i=1}^{t} p(f_i \mid f^{i-1}) p(a_{t+1} \mid a^t, b^t) \prod_{\tilde{b}^t \neq b^t} \sum_{a_{t+1}} p(a_{t+1} \mid f^t(\tilde{b}^{t-1}), \tilde{b}^t)$$

$$
\begin{aligned}
&= \sum_{f^t \in \Upsilon^t(b^{t-1},a^t)} \prod_{i=1}^{t} p(f_i \mid f^{i-1}) p(a_{t+1} \mid a^t, b^t) \\
&= P_{F^T}(\Upsilon^t(b^{t-1}, a^t)) p(a_{t+1} \mid a^t, b^t) \\
&\stackrel{(c)}{=} \vec{p}(a^t \mid b^{t-1}) p(a_{t+1} \mid a^t, b^t) \\
&= \vec{p}(a^{t+1} \mid b^t)
\end{aligned}
$$

where (a) follows because $f^{t+1} \in \Upsilon^{t+1}(b^t, a^{t+1})$, (b) follows from an argument similar to that given above, and (c) follows from the induction hypothesis. $\square$

A function $f_t$ is defined by its graph. In the above construction, equation (8), we have enforced independence across the different $b^{t-1}$. Specifically for $b^{t-1} \neq \bar{b}^{t-1}$ we have

$$P_{F^T}(\Upsilon(b^{t-1}, a_t) \cap \Upsilon(\bar{b}^{t-1}, \bar{a}_t) \mid f^{t-1}) = P_{F^T}(\Upsilon(b^{t-1}, a_t) \mid f^{t-1}) \times P_{F^T}(\Upsilon(\bar{b}^{t-1}, \bar{a}_t) \mid f^{t-1}).$$

We do not need to assume this independence. For example it is known that Gaussian (linear) channel input distributions are optimal for Gaussian channels. For more details see [7], [29], [40], [41]. When dealing with more complicated alphabets one may want the functions $f_t$ to be continuous with respect to the topologies of $\mathcal{A}$ and $\mathcal{B}$. Continuity is trivially satisfied in the finite alphabet case.

Note that it is possible for distinct code-function stochastic kernels to induce the same channel input distribution (almost surely.) In addition, there may be many code-functions stochastic kernels that are good with respect to a given channel input distribution. As an example consider the case when the channel input distribution does not depend on the channel output: $\{p(da_t \mid a^{t-1})\}_{t=1}^{T}$. One choice of $P_{F^T}$ is given in equation (8):

$$p(f_t \mid f^{t-1}) = \prod_{b^{t-1}} p(f_t(b^{t-1}) \mid f^{t-1}(b^{t-2}))$$

Another choice would be to put zero mass on code-functions that depend on feedback (i.e. only use codewords):

$$P_{F^T}(f^T) = \begin{cases} \prod_{t=1}^{T} p(a_t \mid a^{t-1}) & \text{if } f_t(b^{t-1}) = a_t \; \forall b^{t-1}, \; \forall t \\ 0 & \text{else} \end{cases}$$

One can show that this $P_{F^T}(df^T)$ is good with respect to $\{q(da_t \mid a^{t-1})\}$ by checking for each $t$: $P_{F^T}(\Upsilon^t(b^{t-1}, a^t)) = \prod_{i=1}^{t} p(a_t \mid a^{i-1})$.

For memoryless channels we know the optimal channel input distribution is $\{p(da_t)\}_{t=1}^{T}$. Feedback in this case cannot increase capacity but that does not preclude us from using feedback. For example, feedback is known to decrease latency.

## 5.2 Feedback Channel Coding Theorem

Now we can prove the feedback channel coding theorem 5.1. We start with the converse part and then prove the direct part.

**Converse Theorem:** Choose a $(T, M, \epsilon)$ channel code $\{f^T[w]\}_{w=1}^{M}$. Place a prior probability $\frac{1}{M}$ on each code-function $f^T[w]$. By lemma 3.1 and corollary 3.1 this defines consistent measures $Q(df^T, da^T, db^T)$ and $Q(dw, da^T, db^T)$. The following is a generalization of the Verdú-Han converse [33].

**Lemma 5.5** *Every $(T, M, \epsilon)$ channel code satisfies*

$$\epsilon \geq Q\left(\frac{1}{T} \log \frac{Q_{A^T, B^T}(A^T, B^T)}{\vec{Q}_{A^T|B^T} Q_{B^T}(A^T, B^T)} \leq \frac{1}{T} \log M - \gamma\right) - 2^{-\gamma T} \quad \forall \gamma > 0$$

**Proof:** Choose a $\gamma > 0$. Let $D_w \subset \mathcal{B}^T$ be the decoding region for message $w$. The only restriction we place on the decoding regions is that they do not intersect: $D_w \cap D_{\tilde{w}} = \emptyset \ \forall \tilde{w} \neq w$. (This is always true when using a channel decoder: $D_w = \{w : g(b^T) = w\}$.)

Under this restriction on the decoder Verdú and Han show in theorem 4 of [33] that any $(T, M, \epsilon)$ channel code for the channel $\{p(db_t \mid f^t, b^{t-1})\}$ without feedback (see equation (2)) satisfies

$$\epsilon \geq Q\left(\frac{1}{T} \log \frac{Q_{F^T, B^T}(F^T, B^T)}{Q_{F^T} Q_{B^T}(F^T, B^T)} \leq \frac{1}{T} \log M - \gamma\right) - 2^{-\gamma T} \quad \forall \gamma > 0$$

By lemma 5.2 we know that $\frac{Q_{F^T, B^T}(F^T, B^T)}{Q_{F^T} Q_{B^T}(F^T, B^T)} = \frac{Q_{A^T, B^T}(A^T, B^T)}{\vec{Q}_{A^T|B^T} Q_{B^T}(A^T, B^T)}$ holds $Q - a.s.$ $\square$

Note that in the proof of lemma 5.5 the only property of the decoder we used is the restriction that the decoding regions not overlap. Thus the lemma holds independently of the decoder that one uses.

**Theorem 5.2** *The channel capacity $C^O \leq C$.*

**Proof:** Assume towards a contradiction that $C^O > C$. Specifically, assume there exists a sequence of $(T, M_T, \epsilon_T)$ channel codes with $\lim_{T \to \infty} \epsilon_T = 0$ and $\liminf_{T \to \infty} \frac{1}{T} \log M_T > C + 2\gamma$ for some $\gamma > 0$. Then

$$\begin{aligned}\epsilon_T &\geq Q\left(\frac{1}{T} \log \frac{Q_{A^T, B^T}(A^T, B^T)}{\vec{Q}_{A^T|B^T} Q_{B^T}(A^T, B^T)} \leq \frac{1}{T} \log M_T - \gamma\right) - 2^{\gamma T} \\ &\geq Q\left(\frac{1}{T} \log \frac{Q_{A^T, B^T}(A^T, B^T)}{\vec{Q}_{A^T|B^T} Q_{B^T}(A^T, B^T)} \leq C + \gamma\right) - 2^{\gamma T}\end{aligned}$$

where the line follows from lemma 5.5 and the second line holds for all sufficiently large $T$. By the definition of $C$ and for all sufficiently large $T$ the mass below $C + \gamma$ has nonzero probability. Therefore the right hand side in the last inequality is greater than zero. Thus contradicting $\epsilon_T \to 0$. $\square$

**Direct Theorem:** We will prove the direct theorem via a random coding argument. The following is a generalization of Feinstein's lemma [13], [33].

**Lemma 5.6** *Fix a time $T$, an $0 < \epsilon < 1$, and a channel $\{p(db_t \mid b^{t-1}, a^t)\}_{t=1}^T$. Then for all $\gamma > 0$ and any channel input distribution $\{r(da_t \mid a^{t-1}, b^{t-1})\}_{t=1}^T$ there exists a $(T, M, \epsilon)$ channel code that satisfies*

$$\epsilon \leq R_{A^T, B^T}\left(\frac{1}{T}\log \frac{R_{A^T, B^T}(A^T, B^T)}{\vec{R}_{A^T \mid B^T} R_{B^T}(A^T, B^T)} \leq \frac{1}{T}\log M + \gamma\right) + 2^{-\gamma T}$$

*where $R_{A^T, B^T}(da^T, db^T) = \bigotimes_{t=1}^T p(db_t \mid b^{t-1}, a^t) \otimes r(da_t \mid a^{t-1}, b^{t-1})$.*

**Proof:** Let $\{p(df_t \mid f^{t-1})\}_{t=1}^T$ be any sequence of code-function stochastic kernels good with respect to the channel input distribution $\{r(da_t \mid a^{t-1}, b^{t-1})\}_{t=1}^T$. Let $Q(df^T, da^T, db^T)$ be the consistent joint measure determined by this $\{p(df_t \mid f^{t-1})\}_{t=1}^T$ and the channel.

Verdú and Han, theorem 2 of [33], show that for the channel $\{p(db_t \mid f^t, b^{t-1})\}_{t=1}^T$ without feedback and for every $\gamma > 0$ there exists a channel code $(T, M, \epsilon)$ that satisfies:

$$\epsilon \leq Q\left(\frac{1}{T}\log \frac{Q_{F^T, B^T}(F^T, B^T)}{Q_{F^T} Q_{B^T}(F^T, B^T)} \leq \frac{1}{T}\log M + \gamma\right) + 2^{-\gamma T}.$$

Lemma 5.2 shows $\frac{Q_{F^T, B^T}(F^T, B^T)}{Q_{F^T} Q_{B^T}(F^T, B^T)} = \frac{Q_{A^T, B^T}(A^T, B^T)}{\vec{Q}_{A^T \mid B^T} Q_{B^T}(A^T, B^T)}$ holds $Q$–almost surely. Lemma 5.3 shows $Q(a_t \mid a^{t-1}, b^{t-1}) = r(a_t \mid a^{t-1}, b^{t-1})$ $Q$–almost surely. Hence $Q(da^T, db^T) = R_{A^T, B^T}(da^T, db^T)$. □

Recall that the random coding argument underlying this result requires a distribution on channel codes given by randomly drawing $M$ code-functions uniformly from $Q(df^T)$.

**Theorem 5.3** *The channel capacity $C^o \geq C$.*

**Proof:** We follow [33]. Fix an $\epsilon > 0$. We will show that $C$ is an $\epsilon$–achievable rate by demonstrating for every $\delta > 0$ and all sufficiently large $T$ there exists a sequence of $(T, M, 2^{-\frac{T\delta}{4}} + \frac{\epsilon}{2})$ codes with rate $C - \delta \leq \frac{\log M}{T} \leq C - \frac{\delta}{2}$. If in the previous lemma we choose $\gamma = \frac{\delta}{4}$, then we get

$$R_{A^T, B^T}\left(\frac{1}{T}\log \frac{R_{A^T, B^T}(A^T, B^T)}{\vec{R}_{A^T \mid B^T} R_{B^T}(A^T, B^T)} \leq \frac{1}{T}\log M + \frac{\delta}{4}\right)$$
$$\leq \left(\frac{1}{T}\log \frac{R_{A^T, B^T}(A^T, B^T)}{\vec{R}_{A^T \mid B^T} R_{B^T}(A^T, B^T)} \leq C - \frac{\delta}{4}\right)$$
$$\leq \frac{\epsilon}{2}$$

where the second inequality holds for all sufficiently large $T$. To see this note that by the definition of $C$ and $T$ large enough the mass below $C - \frac{\delta}{4}$ has probability zero. □

By combining theorems 5.2 and 5.3 we can conclude theorem 5.1. Specifically $C$ is the feedback channel capacity. It should be clear that if we restrict ourselves to channels without feedback then we recover the original coding theorem by Verdú and Han [33].

**Definition 5.2** *A channel with capacity $C^O$ has a strong converse if for all $\delta > 0$ and every sequence of channel codes, $\{(T, M_T, \epsilon_T)\}$, for which $\liminf \frac{\log M_T}{T} > C^O + \delta$ satisfies $\lim_{T \to \infty} \epsilon_T = 1$.*

Following theorem 7 of [33] we have:

**Proposition 5.1** *A channel has a strong converse if and only if $\sup_{\{\mathcal{D}_T\}_{T=1}^{\infty}} \underline{I}(A \to B) = \sup_{\{\mathcal{D}_T\}_{T=1}^{\infty}} \overline{I}(A \to B)$ and hence $C^O = \lim_{T \to \infty} \frac{1}{T} I(A^T \to B^T)$.*

The latter part follows from theorem 5.1, lemma 4.1, and the finiteness of $\mathcal{B}$.

**Error Exponents:** We can generalize Gallager's [14] error exponent to feedback channels. Specifically, the error exponent for rate $R$ and blocklength $T$ is given by

$$\sup_{\mathcal{D}_T} \max_{0 \le \rho \le 1} \left( -\rho R - \frac{1}{T} \ln \sum_{b^T} \left[ \sum_{a^T} \vec{p}(a^T|b^T) \left\{ \vec{p}(b^T|a^T) \right\}^{\frac{1}{1+\rho}} \right]^{1+\rho} \right).$$

For full details see [29].

## 5.3 General Information Pattern

So far we have assumed that the encoder has access to all the channel outputs $B^{t-1}$. There are many situations, though, where the information pattern [35] at the encoder may be restricted. Let $\mathcal{E}$ be a finite set and let $E_t = \psi_t(B^t)$. Here the measurable functions $\psi_t : \mathcal{B}^t \to \mathcal{E}$ determine the information fed back from the decoder to the encoder. Let $\Psi^T = \{\psi_t\}_{t=1}^T$. In the case of $\Delta$-delayed feedback we have $E_t = \psi_t(B^t) = B_{t-\Delta+1}$. If $\Delta = 1$ then $E_t = \psi_t(B^t) = B_t$ and we are in the situation discussed above. Quantized channel output feedback can be handled by letting the $\{\psi_t\}$ be quantizers. The time ordering is $A_1, B_1, E_1, A_2, B_2, E_2, ..., A_T, B_T, E_T$.

A *channel code-function with information pattern* $\Psi^T$ is a sequence of $T$ deterministic measurable maps $\{f_t\}_{t=1}^T$ such that $f_t : \mathcal{E}^{t-1} \to \mathcal{A}$ taking $e^{t-1} \mapsto a_t$. Denote the set of all code-functions with restricted information pattern $\Psi^T$ by $\mathcal{F}^{T,\Psi} \subseteq \mathcal{F}^T$. The *operational capacity with information pattern* $\Psi^{\infty}$, denoted by $C^{O,\Psi}$, is defined similarly to definition 3.2.

Just as in section 3.1 we can define a joint measure $P(df^T, da^T, db^T, de^T)$ as the interconnection of the code-functions and the channel $\{p(db_t \mid a^t, b^{t-1})\}$. Lemma 3.1 follows as before except that now condition two of consistency requires both $A_t = F^t(E^{t-1}), E_t = \Psi(B^t) \ Q-a.s.$

Define the *channel input distribution with information pattern* $\Psi$ to be a sequence of stochastic kernels $\{p(da_t \mid a^{t-1}, b^{t-1})\}$ with the further condition that for each $t$ the kernel $p(da_t \mid a^{t-1}, b^{t-1}) = p(da_t \mid a^{t-1}, \psi^{t-1}(b^{t-1}))$. Let $\mathcal{D}_T^{\Psi} = \{\{p(da_t \mid a^{t-1}, b^{t-1})\}_{t=1}^T\}$ be the set of all channel input distributions with information pattern $\Psi$. Let

$$C_T^{\Psi} = \sup_{\mathcal{D}_T^{\Psi}} \frac{1}{T} I(A^T \to B^T) \text{ for finite } T \text{ and } C^{\Psi} = \sup_{\{\mathcal{D}_t^{\Psi}\}_{t=1}^{\infty}} \underline{I}(A \to B).$$

For the general information pattern, lemmas 5.1-5.4 and theorems 5.1-5.6 continue to hold with obvious modifications. Hence

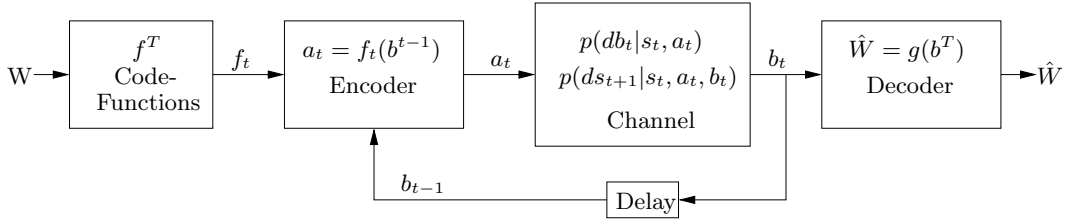

Figure 2: Markov Channel

**Theorem 5.4** *For channels with information pattern $\Psi$ we have $C^{O,\Psi} = C^\Psi$.*

This result holds because the feedback is a causal, deterministic function of the channel outputs. It would be more interesting and practical if the feedback were noisy. This is a more complicated problem as it is related to the problem of channel coding with side-information at the encoder.

## 6 Markov Channels

In this section we formulate the Markov channel feedback capacity problem. As before let $\mathcal{A}, \mathcal{B}$ be spaces with a finite number of elements representing the channel input and channel output, respectively. Furthermore let $\mathcal{S}$ be a state space with a finite number of elements with the counting $\sigma$-algebra. Let $S_t, A_t, B_t$ be measurable random elements taking values in $\mathcal{S}, \mathcal{A}, \mathcal{B}$ respectively. See figure 2. There is a natural time-ordering on the random variables of interest:

$$W,\ S_1,\ A_1,\ B_1,\ S_2,\ \ldots,\ S_t,\ \overbrace{A_t,\ B_t,\ S_{t+1}}^{t\text{--th epoch}},\ \overbrace{A_{t+1},\ B_{t+1},\ S_{t+2}}^{t+1\text{--st epoch}},\ \ldots,\ S_T,\ A_T,\ B_T,\ \hat{W} \quad (9)$$

First, at time 0 a message $W$ is produced and the initial state $S_1$ drawn. The order of events in each of the $T$ epochs is described in (9). At beginning of $t$-th epoch the channel input symbol $A_t$ is placed on the channel by the transmitter, then $B_t$ is observed by the receiver, then the state of the system evolves to $S_{t+1}$, and then finally the receiver feeds back information to the transmitter. At the beginning of the $t+1$ epoch the transmitter uses the feedback information to produce the next channel input symbol $A_{t+1}$. Finally at time $T$, after observing $B_T$, the decoder outputs the reconstructed message $\hat{W}$.

**Definition 6.1** *A* Markov channel *consists of an initial state distribution $p(ds_1)$, the state transition stochastic kernels $\{p(ds_{t+1} \mid s_t, a_t, b_t)\}_{t=1}^{T-1}$, and the channel output stochastic kernels $\{p(db_t \mid s_t, a_t)\}_{t=1}^{T}$. If the stochastic kernel $p(ds_{t+1} \mid s_t, a_t, b_t)$ is independent of $a_t, b_t$ for each $t = 1, ..., T-1$ then we say the channel is a* Markov channel without ISI *(intersymbol interference.)*

Note that we are assuming the kernels $\{p(ds_{t+1} \mid s_t, a_t, b_t)\}$ and $\{p(db_t \mid s_t, a_t)\}$ are stationary (independent of time.)

As before a *channel code-function* is a sequence of $T$ deterministic measurable maps $\{f_t\}_{t=1}^T$ such that $f_t : \mathcal{B}^{t-1} \to \mathcal{A}$ which takes $b^{t-1} \mapsto a_t$. We do not assume, for now, that the state of the channel is observable to the encoder. This will have the effect of restricting ourselves to channel input distributions of the form $\{p(da_t \mid a^{t-1}, b^{t-1})\}$ as opposed to $\{p(da_t \mid s^t, a^{t-1}, b^{t-1})\}$. We do assume that *both* the encoder and the decoder know $p(ds_1)$. In section 8.1 we show how to introduce state feedback.

## 6.1 The Sufficient Statistic $\{\Pi_t\}$

Given a sequence of code-function distributions $\{p(df_t \mid f^{t-1})\}_{t=1}^T$ we can interconnect the Markov channel to the source. Via a straightforward generalization of definition 3.3 and lemma 3.1 one can show there exists a unique consistent measure: $Q(df^T, ds^T, da^T, db^T) = \bigotimes_{t=1}^T p(df_t \mid f^{t-1}) \otimes p(ds_t \mid s_{t-1}, a_{t-1}, b_{t-1}) \otimes \delta_{\{f_t(b^{t-1})\}}(da_t) \otimes p(db_t \mid s_t, a_t)$. Unlike in lemma 3.1 determining the channel without feedback from $\mathcal{F}^T$ to $\mathcal{B}^T$ takes a bit more work. To that end we introduce the sufficient statistics $\{\Pi_t\}$.

Let $\Pi(ds) \in \mathcal{P}(\mathcal{S})$ be an element in the space of probability measures on $\mathcal{S}$. Define a stochastic kernel from $\mathcal{P}(\mathcal{S}) \times \mathcal{A}$ to $\mathcal{S} \times \mathcal{B}$:

$$r(ds, db \mid \pi, a) = p(db \mid s, a) \otimes \pi(ds) \quad (10)$$

The following lemma follows from theorem A.3 in the appendix.

**Lemma 6.1** *There exists a stochastic kernel $r(ds \mid \pi, a, b)$ from $\mathcal{P}(\mathcal{S}) \times \mathcal{A} \times \mathcal{B}$ to $\mathcal{S}$ such that*

$$r(ds, db \mid \pi, a) = r(ds \mid \pi, a, b) \otimes r(db \mid \pi, a)$$

*where $r(db \mid \pi, a)$ is the marginal of $r(ds, db \mid \pi, a)$. Specifically, for each $b$:*

$$r(b \mid \pi, a) = \sum_{\tilde{s}} p(b \mid \tilde{s}, a) \pi(\tilde{s}) \quad (11)$$

The statistic $\pi(ds)$ is often called the *a priori* distribution of the state and $r(ds \mid \pi, a, b)$ the *a posteriori* distribution of the state after observing $a, b$. We recursively define the sufficient statistics $\{\Pi_t\}_{t=1}^T$. Specifically $\Pi_t : \mathcal{A}^{t-1} \times \mathcal{B}^{t-1} \to \mathcal{P}(\mathcal{S})$ defined as follows:

$$\pi_1(ds_1) = p(ds_1) \quad (12)$$

(where $p(ds_1)$ is given in definition 6.1) and for each $a^t, b^t$ and all $s_{t+1}$:

$$\pi_{t+1}[a^t, b^t](s_{t+1}) = \sum_{s_t} p(s_{t+1} \mid s_t, a_t, b_t) r\left(s_t \mid \pi_t[a^{t-1}, b^{t-1}](ds_t), a_t, b_t\right) \quad (13)$$

Equations (12) and (13) are the so-called filtering equations for the state of the channel based on the channel inputs and outputs. Note that equation (13) implies there exists a deterministic, stationary, measurable function $\Phi_\Pi$ such that $\pi_{t+1} = \Phi_\Pi(\pi_t, a_t, b_t)$ for all $t = 1, ..., T-1$. Note the statistic $\Pi_t$ depends on information from *both* the transmitter and the receiver. It can be viewed as the complete system's estimate of the state.

We will now show that the $\{\Pi_t\}$ defined in equations (12) and (13) are consistent with the conditional probabilities $\{Q(ds_t \mid f^t, a^{t-1}, b^{t-1})\}$.

**Lemma 6.2** *We are given a sequence of code-function stochastic kernels $\{p(df_t \mid f^{t-1})\}$, a Markov channel $p(ds_1), \{p(ds_{t+1} \mid s_t, a_t)\}, \{p(db_t \mid s_t, a_t)\}$, and a consistent joint measure $Q(df^T, ds^T, da^T, db^T)$. Then for each $t$ and all $s_t$ we have*

$$Q(s_t \mid f^t, a^t, b^{t-1}) = \pi_t[a^{t-1}, b^{t-1}](s_t) \tag{14}$$

*for $Q$ almost all $f^t, a^t, b^{t-1}$.*

**Proof:** We will prove equation (14) by induction. For $t = 1$ and all $s_1$ we have $\pi_1(s_1)Q(f_1, a_1)$ $= p(s_1)p(f_1)\delta_{\{f_1\}}(a_1) = Q(f_1, s_1, a_1)$. Now for $t+1$ and all $s_{t+1}$ we have

$$\pi_{t+1}[a^t, b^t](s_{t+1}) \; Q(f^{t+1}, a^{t+1}, b^t)$$
$$= \pi_{t+1}[a^t, b^t](s_{t+1}) \sum_{s_t} Q(f^{t+1}, s_t, a^{t+1}, b^t)$$

$$\stackrel{(a)}{=} \left( \sum_{\tilde{s}_t} p(s_{t+1} \mid \tilde{s}_t, a_t, b_t) r\left(\tilde{s}_t \mid \pi_t[a^{t-1}, b^{t-1}](ds_t), a_t, b_t\right) \right)$$
$$\times \left( \sum_{s_t} \delta_{\{f_{t+1}(b^t)\}}(a_{t+1}) \; p(f_{t+1} \mid f^t) \; p(b_t \mid s_t, a_t) \; \delta_{\{f_t(b^{t-1})\}}(a_t)\pi[a^{t-1}, b^{t-1}](s_t) \; Q(f^t, a^{t-1}, b^{t-1}) \right)$$

$$= \sum_{\tilde{s}_t} p(s_{t+1} \mid \tilde{s}_t, a_t, b_t) \left( r\left(\tilde{s}_t \mid \pi_t[a^{t-1}, b^{t-1}](ds_t), a_t, b_t\right) \sum_{s_t} p(b_t \mid s_t, a_t) \; \pi[a^{t-1}, b^{t-1}](s_t) \right)$$
$$\times \delta_{\{f_{t+1}(b^t)\}}(a_{t+1}) p(f_{t+1} \mid f^t) \delta_{\{f_t(b^{t-1})\}}(a_t) \; Q(f^t, a^{t-1}, b^{t-1})$$

$$\stackrel{(b)}{=} \sum_{\tilde{s}_t} p(s_{t+1} \mid \tilde{s}_t, a_t, b_t) \left( p(b_t \mid \tilde{s}_t, a_t) \; \pi[a^{t-1}, b^{t-1}](\tilde{s}_t) \right)$$
$$\times \delta_{f_{t+1}(b^t)}(a_{t+1}) p(f_{t+1} \mid f^t) \; \delta_{\{f_t(b^{t-1})\}}(a_t) \; Q(f^t, a^{t-1}, b^{t-1})$$

$$\stackrel{(c)}{=} \sum_{\tilde{s}_t} \delta_{\{f_{t+1}(b^t)\}}(a_{t+1}) \; p(s_{t+1} \mid \tilde{s}_t, a_t, b_t) \; p(f_{t+1} \mid f^t) \; p(b_t \mid \tilde{s}_t, a_t) \; \delta_{\{f_t(b^{t-1})\}}(a_t) \; Q(f^t, \tilde{s}_t, a^{t-1}, b^{t-1})$$

$$= \sum_{\tilde{s}_t} Q(f^{t+1}, \tilde{s}_t, s_{t+1}, a^{t+1}, b^t)$$

$$= Q(f^{t+1}, s_{t+1}, a^{t+1}, b^t)$$

where (a) follows from the definition of $\Pi_t$ and the induction hypothesis. Line (b) follows from lemma 6.1 and (c) is another application of the induction hypothesis. $\square$

Note that equation (14) states that the conditional probability $Q(ds_t \mid f^t, a^t, b^{t-1})$ does not depend on $f^t$ almost surely. In addition the filtering equations (12) and (13) are defined independently of the code-function distributions (or equivalently the channel input distributions). This is related to Witsenhausen's work on policy independence [36]. Finally observe that equation (14) and the fact that $\Pi_t$ is a function of $A^{t-1}, B^{t-1}$ imply that $S_t - \Pi_t - (F^t, A^t, B^{t-1})$ forms a Markov chain under any consistent measure $Q$.

## 6.2 Markov Channel Coding Theorem

We are now in a position to describe the "$\mathcal{F}^T - \mathcal{B}^T$" channel in terms of the underlying Markov channel. We then prove the Markov channel coding theorem.

**Lemma 6.3** We are given a sequence of code-function stochastic kernels $\{p(df_t \mid f^{t-1})\}$; a Markov channel $p(ds_1), \{p(ds_{t+1} \mid s_t, a_t, b_t)\}, \{p(db_t \mid s_t, a_t)\}$; and a consistent joint measure $Q(df^T, ds^T, da^T, db^T)$. Then for each $t$ and all $b_t$ we have

$$Q(b_t \mid f^t, a^t, b^{t-1}) = r\left(b_t \mid \pi_t[a^{t-1}, b^{t-1}](ds_t), a_t\right) \tag{15}$$

for $Q$ almost all $f^t, a^t, b^{t-1}$. Where $r(db \mid \pi, a)$ was defined in equation (11).

**Proof:** For each $t$ note that

$$\begin{aligned} Q(f^t, a^t, b^t) &= \sum_{s_t} Q(f^t, s_t, a^t, b^t) \\ &= \sum_{s_t} p(b_t \mid s_t, a_t)\, Q(f^t, s_t, a^t, b^{t-1}) \\ &\stackrel{(a)}{=} \sum_{s_t} p(b_t \mid s_t, a_t)\, \pi_t[a^{t-1}, b^{t-1}](s_t)\, Q(f^t, a^t, b^{t-1}) \\ &= r(b_t \mid \pi_t[a^{t-1}, b^{t-1}](ds_t), a_t)\, Q(f^t, a^t, b^{t-1}) \end{aligned}$$

where (a) follows from lemma 6.2. □

The previous lemma shows that $B - (\Pi_t, A_t) - (F^t, A^{t-1}, B^{t-1})$ forms a Markov chain under $Q$.

**Corollary 6.1** We are given a sequence of code-function stochastic kernels $\{p(df_t \mid f^{t-1})\}$; a Markov channel $p(ds_1), \{p(ds_{t+1} \mid s_t, a_t, b_t)\}, \{p(db_t \mid s_t, a_t)\}$; and a consistent joint measure $Q(df^T, ds^T, da^T, db^T)$. Then for each $t$ and all $b_t$ we have

$$Q(b_t \mid f^t, b^{t-1}) = r\left(b_t \mid \pi_t[f^{t-1}(b^{t-2}), b^{t-1}](ds_t), f_t(b^{t-1})\right) \tag{16}$$

for $Q$ almost all $f^t, b^{t-1}$.

**Proof:** For each $t$ note that

$$\begin{aligned} Q(f^t, b^t) &= \sum_{a^t} Q(f^t, a^t, b^t) \\ &= \sum_{a^t} r(b_t \mid \pi_t[a^{t-1}, b^{t-1}](ds_t), a_t)\, Q(f^t, a^t, b^{t-1}) \\ &= r(b_t \mid \pi_t[f^{t-1}(b^{t-2}), b^{t-1}](ds_t), f_t(b^{t-1}))\, Q(f^t, b^{t-1}) \end{aligned}$$

where the second line follows from lemma 6.3. □

The corollary shows that we can convert a Markov channel into a channel of the general form considered in sections 3-5. Hence we can define the operational channel capacity, $C^O$, for the Markov channel with feedback in exactly the same way we did in definition 4.3. We can also use the same definitions of capacity, $C$, as before. Thus we can directly apply theorem 5.1 and its generalization, theorem 5.4, to prove:

**Theorem 6.1** For Markov channels we have $C^O = C$. For Markov channels with information pattern $\Psi$ we have $C^{O,\Psi} = C^\Psi$.

We end this section by noting that the use of $\{\Pi_t\}$ can simplify the form of the directed information and the choice of the channel input distribution.

**Lemma 6.4** *For a Markov channel $I(F^T \to B^T) = I(A^T \to B^T) = \sum_{t=1}^T I(A_t, \Pi_t; B_t \mid B^{t-1})$.*

**Proof:** The first equality follows from lemma 5.2. The second equality follows from noting that $I(A^T \to B^T) = \sum_{t=1}^T I(A^t; B_t \mid B^{t-1})$. For $t=1$ we know $\Pi_1(ds_1) = p(ds_1)$ is a fixed, non-random, measure known to both the transmitter and receiver. Hence $I(A_1; B_1) = I(A_1, \Pi_1; B_1)$. For $t > 1$ we have

$$\begin{aligned} I(A^t; B_t \mid B^{t-1}) &= I(A_t, \Pi_t; B_t \mid B^{t-1}) + I(A^{t-1}; B_t \mid \Pi_t, A_t, B^{t-1}) - I(\Pi_t; B_t \mid A^t, B^{t-1}) \\ &= I(A_t, \Pi_t; B_t \mid B^{t-1}) \end{aligned}$$

where $I(\Pi_t; B_t \mid A^t, B^{t-1}) = 0$ because $\Pi_t$ is a function of $A^{t-1}, B^{t-1}$. Lemma 6.3 implies $(A^{t-1}, B^{t-1}) - (\Pi_t, A_t) - B_t$ is a Markov chain hence $I(A^{t-1}; B_t \mid \Pi_t, A_t, B^{t-1}) = 0$. □

Note that we can view both $(A_t, \Pi_t)$ as the input to the channel. This makes sense because the decoder needs information about the encoder's estimate of the state given by $\Pi_t$. The next lemma shows us that we simplify the form of the channel input distribution.

**Lemma 6.5** *Given a Markov channel $p(ds_1), \{p(ds_{t+1} \mid s_t, a_t, b_t)\}, \{p(db_t \mid s_t, a_t)\}$, and a channel input distribution $\{q(da_t \mid a^{t-1}, b^{t-1})\}$ with resulting joint measure $Q(ds^T, da^T, db^T)$ there exists another channel input distribution of the form $\{r(da_t \mid \pi_t, b^{t-1})\}$ with resulting joint measure $R(ds^T, da^T, db^T)$ such that for each $t$ we have[2]*

$$R(d\pi_t, da_t, db^t) = Q(d\pi_t, da_t, db^t)$$

*and hence $I_R(A_t, \Pi_t; B_t \mid B^{t-1}) = I_Q(A_t, \Pi_t; B_t \mid B^{t-1})$.*

**Proof:** From lemmas 6.2 and 6.3 and equation (13) we know

$$Q(d\pi^T, da^T, db^T) = \bigotimes_{t=1}^T r(db_t \mid \pi_t, a_t) \otimes q(da_t \mid a^{t-1}, b^{t-1}) \otimes \delta_{\{\Phi_\Pi(\pi_{t-1}, a_{t-1}, b_{t-1})\}}(d\pi_t) \quad (17)$$

where, as an abuse of notation, let $\delta_{\{\Phi_\Pi(\pi_0, a_0, b_0)\}}(d\pi_1) = \delta_{\{p(ds_1)\}}(d\pi_1)$. For each $t$ define the stochastic kernel $r(da_t \mid \pi_t, b^{t-1})$ to be a version of the conditional distribution $Q(da_t \mid \pi_t, b^{t-1})$ (see theorem A.3 in the appendix.)

We proceed by induction. For $t=1$ we know $\pi_1(ds_1) = p(ds_1)$. For any Borel measurable set $\Omega \subset \mathcal{P}(\mathcal{S})$, $a_1, b_1$ we have

$$R(\Omega, a_1, b_1) = \int_\Omega r(b_1 \mid \pi_1, a_1) r(a_1 \mid \pi_1) \delta_{\{p(ds_1)\}}(d\pi_1) = \int_\Omega r(b_1 \mid \pi_1, a_1) Q(d\pi_1, a_1) = Q(\Omega, a_1, b_1).$$

---

[2]For any Borel measurable $\Omega \subset \mathcal{P}(\mathcal{S})$ let $Q(\pi_t \in \Omega,\ a_t = \bar{a}_t,\ b^t = \bar{b}^t) = Q\left(\{(a^t, b^t)\ :\ a_t = \bar{a}_t,\ b^t = \bar{b}^t,\ \pi_t[a^{t-1}, b^{t-1}] \in \Omega\}\right).$

Now for $t+1$ and any Borel measurable set $\Omega \subset \mathcal{P}(\mathcal{S})$, $a_{t+1}, b^{t+1}$ we have

$$\begin{aligned}
&R(\pi_{t+1} \in \Omega, a_{t+1}, b^{t+1}) \\
&= \sum_{a_t} \int_\Omega \int_{\mathcal{P}(\mathcal{S})} R(d\pi_t, d\pi_{t+1}, a_t, a_{t+1}, b^{t+1}) \\
&= \sum_{a_t} \int_\Omega \int_{\mathcal{P}(\mathcal{S})} r(b_{t+1} \mid \pi_{t+1}, a_{t+1}) r(a_{t+1} \mid \pi_{t+1}, b^t) \delta_{\{\Phi_\Pi(\pi_t, a_t, b_t)\}}(d\pi_{t+1}) R(d\pi_t, a_t, b^t) \\
&\stackrel{(a)}{=} \sum_{a_t} \int_\Omega \int_{\mathcal{P}(\mathcal{S})} r(b_{t+1} \mid \pi_{t+1}, a_{t+1}) r(a_{t+1} \mid \pi_{t+1}, b^t) \delta_{\{\Phi_\Pi(\pi_t, a_t, b_t)\}}(d\pi_{t+1}) Q(d\pi_t, a_t, b^t) \\
&= \sum_{a_t} \int_\Omega \int_{\mathcal{P}(\mathcal{S})} r(b_{t+1} \mid \pi_{t+1}, a_{t+1}) r(a_{t+1} \mid \pi_{t+1}, b^t) Q(d\pi_t, d\pi_{t+1}, a_t, b^t) \\
&= \int_\Omega r(b_{t+1} \mid \pi_{t+1}, a_{t+1}) r(a_{t+1} \mid \pi_{t+1}, b^t) Q(d\pi_{t+1}, b^t) \\
&\stackrel{(b)}{=} \int_\Omega r(b_{t+1} \mid \pi_{t+1}, a_{t+1}) Q(d\pi_{t+1}, a_{t+1}, b^t) \\
&= Q(\pi_{t+1} \in \Omega, a_{t+1}, b^{t+1})
\end{aligned}$$

where (a) follows from the induction hypothesis and (b) follows by the construction of $r(da_{t+1} \mid \pi_{t+1}, b^t)$. $\square$

The lemma states that we can without loss of generality restrict ourselves to channel input distributions of the form $\{q(da_t \mid \pi_t, b^{t-1})\}$. Note that the dependence on $a^{t-1}$ appears only through $\pi_t[a^{t-1}, b^{t-1}](ds_t)$. If $\pi_t[a^{t-1}, b^{t-1}]$ is not a function of $a^{t-1}$ then the distribution of $a_t$ will depend only on the feedback $b^{t-1}$. We discuss when this happens in section 8.

In summary, we have shown that any Markov channel, $p(ds_1)$, $\{p(ds_{t+1} \mid s_t, a_t, b_t)\}$, $\{p(db_t \mid s_t, a_t)\}$ can be converted into another Markov channel with initial state $\Pi_1(d\pi_1) = \delta_{\{p(ds_1)\}}(d\pi_1)$, deterministic state transitions $\Pi_{t+1} = \Phi_\Pi(\Pi_t, A_t, B_t)$, and channel output stochastic kernels $\{r(db_t \mid \pi_t, a_t)\}$. We call this the *canonical* Markov channel associated with the original Markov channel. Thus the problem of determining the capacity of a Markov channel with state space $\mathcal{S}$ has been reduced to determining the capacity of the canonical Markov channel. The latter Markov channel has state space $\mathcal{P}(\mathcal{S})$ and state computable from the channel inputs and outputs.

Note that even if the original Markov channel does not have ISI it is typically the case that the canonical Markov channel will have ISI. This is because the choice of channel input can help the decoder identify the channel. This property is called *dual control* in the stochastic control literature [3].

# 7 The MDP Formulation

Our goal in this section is to formulate the following optimization problem for Markov channels with feedback as an infinite horizon average cost problem:

$$\sup_{\mathcal{D}_\infty} \liminf_{T\to\infty} \frac{1}{T} I(A^T \to B^T) = \sup_{\mathcal{D}_\infty} \liminf_{T\to\infty} \frac{1}{T} \sum_{t=1}^T I(A_t, \Pi_t; B_t \mid B^{t-1}). \tag{18}$$

We first give a high-level discussion of the issues, then we formulate the optimization problem in equation (18) as a partially observed Markov decision problem (POMDP), convert the POMDP to a fully observed MDP, and provide the average cost optimality equation (ACOE).

Before proceeding the reader may notice that the optimization in equation (18) is different than the one given after definition 4.4: $C = \sup_{\{\mathcal{D}_T\}_{T=1}^\infty} \underline{I}(A \to B)$. In the course of this section it will be shown that the two optimizations are equivalent. That one can without loss of generality restrict the optimization to $D_\infty$ instead of $\{\mathcal{D}_T\}_{T=1}^\infty$ is a consequence of Bellman's *principle of optimality*. In addition conditions will be given such that under the optimal channel input distribution we have $\liminf_{T\to\infty} \frac{1}{T} I(A^T \to B^T) = \underline{I}(A \to B)$.

To compute $I(A_t, \Pi_t; B_t \mid B^{t-1})$ we need to know the measure:

$$Q(d\pi_t, da_t, db^t) = r(db_t \mid \pi_t, a_t) \otimes q(da_t \mid \pi_t, b^{t-1}) \otimes Q(d\pi_t, db^{t-1}). \tag{19}$$

By lemma 6.5 we know that we can without loss of generality restrict ourselves to channel input distributions of the form $\{q(da_t \mid \pi_t, b^{t-1})\}$.

To formulate the optimization in (18) as a stochastic control problem we need to specify the state space, the control actions, and the running cost. On first glance it may appear that the encoder should choose control actions of the form $u_t(da_t)$ based on the information $(\pi_t[a^{t-1}, b^{t-1}], b^{t-1})$. Unfortunately one cannot write the running cost in terms of $u_t(da_t)$. To see this observe that the argument under the expectation in $I(A_t, \Pi_t; B_t \mid B^{t-1}) = E\left[\log \frac{r(B_t \mid \Pi_t, A_t)}{Q(B_t \mid B^{t-1})}\right]$ can be written as

$$\log \frac{r(b_t \mid a_t, \pi_t)}{Q(b_t \mid b^{t-1})} = \log \frac{r(b_t \mid a_t, \pi_t)}{\int \int r(b_t \mid \tilde{\pi}_t, \tilde{a}_t) Q(d\tilde{\pi}_t, d\tilde{a}_t \mid b^{t-1})} \quad Q-\text{almost all } \pi_t, a_t, b^t \tag{20}$$

This depends on $Q(d\pi_t, da_t \mid b^{t-1})$ and not $Q(da_t \mid \pi_t, b^{t-1})$.

This suggests that the control actions should be stochastic kernels of the form $u_t(da_t \mid \pi_t)$. This too is problematic. Note that we are interested in an optimization given in equation (18) and hence would like for there to be a topology on the space of stochastic kernels of the form $u_t(da_t \mid \pi_t)$. In some cases there is a natural parameterization of this space. For example, for Gaussian channels it is known that the optimal input distribution is linear and can be parameterized by its coefficients [7], [29], [40], [41]. But in general there is no, at least to the author's knowledge, natural topology on the space of stochastic kernels. Hence we will choose control actions of the form $u_t(d\pi_t, da_t)$. The next section formalizes the stochastic control problem with this choice of control action.

## 7.1 Partially Observed Markov Decision Problem

Here we first describe the components of the POMDP formulation. In the next section we show the equivalence of the POMDP formulation to the optimization (18).

Consider the control action $u(d\pi, da)$ in the control space $\mathcal{U} = \mathcal{P}(\mathcal{P}(\mathcal{S}) \times \mathcal{A})$. The space $\mathcal{U}$ is a Polish space (i.e. a complete, separable metric space) equipped with the topology of weak convergence.

The state at time $t > 1$ is $X_t = (\Pi_{t-1}, A_{t-1}, B_{t-1}) \in \mathcal{P}(\mathcal{S}) \times \mathcal{A} \times \mathcal{B}$ and $X_1 = \emptyset$. The dynamics are given as:

$$r(dx_{t+1} \mid x_t, u_t) = r(db_t \mid \pi_t, a_t) \otimes u_t(d\pi_t, da_t) \tag{21}$$

Note that they dynamics depend only on $u_t$. The observation at time $t > 1$ is given by $Y_t = B_{t-1}$ and $Y_1 = \emptyset$. Note that $Y_t$ is a deterministic function of $X_t$.

As discussed, one of the main difficulties in formulating (18) as a POMDP has to do with the form of the cost (20). The cost at time $t$ is given as

$$c(x_t, u_t, x_{t+1}) = \begin{cases} \log \frac{r(b_t \mid \pi_t, a_t)}{\int r(b_t \mid \tilde{\pi}_t, \tilde{a}_t) u_t(d\tilde{\pi}_t, d\tilde{a}_t)} & \text{if } \int r(b_t \mid \tilde{\pi}_t, \tilde{a}_t) u_t(d\tilde{\pi}_t, d\tilde{a}_t) > 0 \\ 0 & \text{else} \end{cases} \tag{22}$$

Note that the cost is just a function of $u_t, x_{t+1}$.

The information pattern at the controller at time $t$ is $(Y^t, U^{t-1}) = (B^{t-1}, U^{t-1}) \in \mathcal{B}^{t-1} \times \mathcal{U}^{t-1}$. The policy at time $t$ is a stochastic kernel $\mu_t(du_t \mid b^{t-1}, u^{t-1})$ from $\mathcal{B}^{t-1} \times \mathcal{U}^{t-1}$ to $\mathcal{U}$. A policy $\{\mu_t\}$ is said to be a *deterministic* policy if for each $t$ and all $(b^{t-1}, u^{t-1})$ the stochastic kernel $\mu_t(du_t \mid b^{t-1}, u^{t-1})$ assigns mass one to only one point in $\mathcal{U}$. In this case we will abuse notation and write $u_t = \mu_t[b^{t-1}]$. Technically, we should explicitly include $p(ds_1)$ and the other channel parameters in the information pattern. But because the channel parameters are fixed throughout and to reduce notation we shall not explicitly mention the control policy's dependence on them.

The time-order of events is the usual one for POMDPs: $X_1, Y_1, U_1, X_2, Y_2, U_2...$. For a given policy $\{\mu_t\}$ the resulting joint measure is

$$R^\mu(du^T, d\pi^T, da^T, db^T) = \bigotimes_{t=1}^{T} r(db_t \mid \pi_t, a_t) \otimes u_t(d\pi_t, da_t) \otimes \mu_t(du_t \mid u^{t-1}, b^{t-1}) \tag{23}$$

where we have used equation (21). Note that this $R$ measure is not the same as the one used in equation (17) of lemma 6.5. Compare the differences between the $R$ measure given in (23) and the $Q$ measure given in equation (17). The next two sections discuss the relation between these two different measures.

## 7.2 The Sufficient Statistic $\{\Gamma_t\}$ and the Control Constraints

The dynamics given in equation (21), the control policy $\{\mu_t\}$, and the running cost given in equation (22) are not enough to specify the optimization in equation (18). In particular, in the original optimization $\{\Pi_t\}$ is determined by (13). Whereas in the POMDP optimization the $\{\Pi_t\}$ are determined by the policy $\{\mu_t\}$. We need to insure that the $\{\Pi_t\}$ play similar roles in both cases. To this end we we will next define appropriate control constraints.

Equation (21) states $r(d\pi, da, db \mid u) = r(db \mid \pi, a) \otimes u(d\pi, da)$. The following lemma follows from theorem A.3 in the appendix.

**Lemma 7.1** *There exists a stochastic kernel $r(d\pi, da \mid u, b)$ from $\mathcal{U} \times \mathcal{B}$ to $\mathcal{P}(\mathcal{S}) \times \mathcal{A}$ such that $r(d\pi, da, db \mid u) = r(d\pi, da \mid u, b) \otimes r(db \mid u)$ where $r(db \mid u)$ is the marginal of $r(d\pi, da, db \mid u)$.*

We now define the statistics $\{\Gamma_t\} \in \mathcal{P}(\mathcal{P}(\mathcal{S}))$. This is the space of probability measures on probability measures on $\mathcal{S}$. Specifically $\Gamma_t : \mathcal{U}^{t-1} \times \mathcal{B}^{t-1} \to \mathcal{P}(\mathcal{P}(\mathcal{S}))$ is defined as follows. For $t = 1$ let

$$\gamma_1(d\pi_1) = \delta_{\{p(ds_1)\}}(d\pi_1) \tag{24}$$

and for $t > 1$ and each $u^{t-1}, b^{t-1}$ and all Borel measurable $\Omega \subset \mathcal{P}(\mathcal{S})$:

$$\gamma_t[u^{t-1}, b^{t-1}](\Omega) = \int \int \{\Phi_\Pi(\pi_{t-1}, a_{t-1}, b_{t-1}) \in \Omega\} r(d\pi_{t-1}, da_{t-1} \mid u_{t-1}, b_{t-1}). \tag{25}$$

Here $\{\cdot\}$ corresponds to the indicator function. Note that for $t > 1$, $\gamma_t[u^{t-1}, b^{t-1}](d\pi_t)$ depends only on $u_{t-1}, b_{t-1}$. Sometimes, for $t > 1$, we will just write $\gamma_t[u_{t-1}, b_{t-1}](d\pi_t)$.

Equation (25) implies there exists a deterministic, stationary, measurable function $\Phi_\Gamma$ such that $\gamma_{t+1} = \Phi_\Gamma(u_t, b_t)$ for all $t = 1, ..., T - 1$. Note that because of feedback the statistic $\Gamma_t$ can be computed at both the transmitter and the receiver. It can be viewed as the receiver's estimate of the transmitter's estimate of the state of the channel.

We now define the control constraints. Let

$$\mathcal{U}(\gamma) = \{u(d\pi, da) \; : \; u(d\pi, da) \in \mathcal{U}, \; u(d\pi) = \gamma(d\pi)\}. \tag{26}$$

Note that for each $\gamma \in \mathcal{P}(\mathcal{P}(\mathcal{S}))$ the set $\mathcal{U}(\gamma)$ is compact. For each $t$ and $(u^{t-1}, b^{t-1})$ the control constraint $\mathcal{U}_t(\cdot) \subset \mathcal{U}$ is defined as:

$$\mathcal{U}_t(u^{t-1}, b^{t-1}) = \mathcal{U}(\gamma_t[u^{t-1}, b^{t-1}]). \tag{27}$$

For each $t$ the policy $\mu_t$ will enforce the control constraint. Specifically for all $(u^{t-1}, b^{t-1})$

$$\mu_t\left(\{u_t \in \mathcal{U}_t\left(\gamma_t[u^{t-1}, b^{t-1}]\right)\} \mid u^{t-1}, b^{t-1}\right) = 1. \tag{28}$$

The next lemma shows that the $\{\Gamma_t\}$ are consistent with the conditional probabilities $R^\mu(d\pi_t \mid u^{t-1}, b^{t-1})$.

**Lemma 7.2** *We are given $p(ds_1)$, the dynamics (21), and a policy $\{\mu_t\}$ satisfying the control constraint (28) with resulting measure $R^\mu(du^T, d\pi^T, da^T, db^T)$. Then for each $t$ we have:*

$$R^\mu(d\pi_t \mid u^{t-1}, b^{t-1}) = \gamma_t[u^{t-1}, b^{t-1}](d\pi_t) \tag{29}$$

*for $R^\mu$ almost all $u^{t-1}, b^{t-1}$.*

**Proof:** Fix a Borel measurable set $\Omega \subset \mathcal{P}(\mathcal{S})$. For any $t$, any Borel measurable sets $\Theta_k \subset \mathcal{U}$, $k = 1, ..., t-1$ and any $b^{t-1}$ we have

$$\begin{aligned}
R^\mu(\Omega, \Theta^{t-1}, b^{t-1}) &= \int_\mathcal{U} \int_{\Theta^{t-1}} \int_\Omega R^\mu(d\pi_t, du^{t-1}, du_t, b^{t-1}) \\
&= \int_\mathcal{U} \int_{\Theta^{t-1}} u_t(\Omega, \mathcal{A}) \mu_t(du_t \mid u^{t-1}, b^{t-1}) R^\mu(du^{t-1}, b^{t-1}) \\
&= \int_{\Theta^{t-1}} \left( \int_\mathcal{U} u_t(\Omega, \mathcal{A}) \mu_t(du_t \mid u^{t-1}, b^{t-1}) \right) R^\mu(du^{t-1}, b^{t-1}) \\
&= \int_{\Theta^{t-1}} \gamma_t[u^{t-1}, b^{t-1}](\Omega) R^\mu(du^{t-1}, b^{t-1})
\end{aligned}$$

where the last equality follows because the control policy $\mu_t$ satisfies the control constraint given in equation (28). $\square$

Equations (29) and (25) show the conditional probability $R^\mu(d\pi_t \mid u^{t-1}, b^{t-1})$ does not depend on the policy $\mu$ and $u^{t-2}, b^{t-2}$ almost surely. See comments after lemma 6.2.

We can simplify the form of the cost, in the standard way, by computing the expectation over the next state. For each $t$ define:

$$\begin{aligned}
\bar{c}(u_t) &= E_{R^\mu}[c(X_t, U_t, X_{t+1}) \mid u_t] \\
&\stackrel{(a)}{=} \int r(db_t \mid \pi_t, a_t) u_t(d\pi_t, da_t) \log \frac{r(b_t \mid \pi_t, a_t)}{\int r(b_t \mid \tilde{\pi}_t, \tilde{a}_t) u_t(d\tilde{\pi}_t, d\tilde{a}_t)}
\end{aligned} \quad (30)$$

where (a) follows from equation (22) and the fact that $c_t$ does not depend on $x_t$.

In summary, we have formulated an average cost, infinite horizon, POMDP:

$$\sup_{\{\mu_t\}} \liminf_{T \to \infty} \frac{1}{T} \sum_{t=1}^T E_{R^\mu}[c(X_t, U_t, X_{t+1})] = \sup_{\{\mu_t\}} \liminf_{T \to \infty} \frac{1}{T} \sum_{t=1}^T E_{R^\mu}[\bar{c}(U_t)] \quad (31)$$

with dynamics given by (21) and costs given by (22). The supremization is over all policies that satisfy the control constraint (28). In the next section we show that the optimization in (31) is equivalent to the optimization in (18).

### 7.3 Equivalence of the Optimization Problems

We now show the equivalence of the optimization problem posed in equation (18) and that posed in (31). As discussed at the end of section 7.1 the measures $Q$ and $R^\mu$ can be different. By equivalence we mean that for any choice of channel input distribution $\{q(da_t \mid \pi_t, b^{t-1})\}$ with resulting joint measure $Q(ds^T, da^T, db^T)$ we can find a control policy $\{\mu_t\}$ satisfying (28) with resulting joint measure $R^\mu(du^T, d\pi^T, da^T, db^T)$ such that for each $t$:

$$Q(d\pi_t, da_t, db^t) = R^\mu(d\pi_t, da_t, db^t). \quad (32)$$

Vice-versa, given any policy $\{\mu_t\}$ satisfying (28) we can find a channel input distribution $\{q(da_t \mid \pi_t, b^{t-1})\}$ such that the above marginals are equal. This equivalence will imply that the optimal costs for the two problems are the same and the optimal channel input distribution is related to the optimal policy.

**Lemma 7.3** *For every channel input distribution $\{q(da_t \mid \pi_t, b^{t-1})\}$ with resulting joint measure $Q(ds^T, da^T, db^T)$ there exists a deterministic policy $\{\mu_t\}$ satisfying (28) with resulting joint measure $R^\mu(du_t, d\pi^T, da^T, db^T)$ such that for each $t$: $R^\mu(d\pi_t, da_t, db^t) = Q(d\pi_t, da_t, db^t)$.*

**Proof:** For each $t$ choose a *deterministic policy* that satisfies:

$$\mu_t[b^{t-1}](d\pi_t, da_t) = Q(d\pi_t, da_t \mid b^{t-1})$$

for $Q$ almost all $b^{t-1}$. We proceed by induction. For $t = 1$ we have $R^\mu(d\pi_1, da_1, db_1) = r(b_1 \mid \pi_1, a_1) \otimes \mu_1[p(ds_1)](d\pi_1, da_1) = r(b_1 \mid \pi_1, a_1) \otimes Q(d\pi_1, da_1) = Q(d\pi_1, a_1, b_1)$. For $t + 1$ we have for any Borel measurable $\Omega \subset \mathcal{P}(\mathcal{S})$ and all $a_{t+1}, b^{t+1}$:

$$\begin{aligned} R^\mu(\Omega, a_{t+1}, b^{t+1}) &= \int_\Omega r(b_{t+1} \mid \pi_{t+1}, a_{t+1}) \mu_{t+1}[b^t](d\pi_{t+1}, a_{t+1}) \; R^\mu(b^t) \\ &\stackrel{(a)}{=} \int_\Omega r(b_{t+1} \mid \pi_{t+1}, a_{t+1}) Q(d\pi_{t+1}, da_{t+1} \mid b^t) Q(b^t) \\ &= Q(\Omega, a_{t+1}, b^{t+1}) \end{aligned}$$

where (a) follows from the induction hypothesis and the our choice of $\mu_{t+1}$.

Now we show the policy $\{\mu_t\}$ satisfies the control constraint (28). For $t = 1$ we have $\mu_1(d\pi_1) = Q(d\pi_1) = \gamma_1(d\pi_1)$. For $t > 1$ we have for any Borel measurable $\Omega \subset \mathcal{P}(\mathcal{S})$ and all $b^{t-1}$:

$$\mu_t[b^{t-1}](\Omega) R^\mu(b^{t-1})$$

$$\stackrel{(a)}{=} Q(\Omega, b^{t-1})$$

$$\stackrel{(b)}{=} \iint \{\Phi_\Pi(\pi_{t-1}, a_{t-1}, b_{t-1}) \in \Omega\} \; r(b_{t-1} \mid \pi_{t-1}, a_{t-1}) \; \mu_t[b^{t-2}](d\pi_{t-1}, da_{t-1}) \; Q(b^{t-2})$$

$$\stackrel{(c)}{=} \iint \{\Phi_\Pi(\pi_{t-1}, a_{t-1}, b_{t-1}) \in \Omega\} \; r(d\pi_{t-1}, da_{t-1} \mid \mu_t[b^{t-2}], b_{t-1}) \; r(b_{t-1} \mid \mu_t[b^{t-2}]) \; R^\mu(b^{t-2})$$

$$\stackrel{(d)}{=} \gamma_t[\mu_{t-1}[b^{t-2}], b_{t-1}](\Omega) \; R^\mu(b^{t-1})$$

where (a) follows from the first part and the choice of control; (b) follows from our choice of control; (c) follows from the first part and lemma 7.1; and (d) follows from equation (25). Finally, altering $\mu_t$ on a set of measure zero if necessary we can insure that for each $t$ the deterministic policy $\mu_t$ will enforce the control constraint. Specifically for each $b^{t-1}$ we have $\mu_t[b^{t-1}] \in \mathcal{U}(\gamma_t[\mu_{t-1}[b^{t-2}], b_{t-1}])$. $\square$

**Lemma 7.4** *For every policy $\{\mu_t\}$ satisfying (28) with resulting joint measure $R^\mu(du^T, d\pi^T, da^T, db^T)$ there exists a channel input distribution $\{q(da_t \mid \pi_t, b^{t-1})\}$ with resulting joint measure $Q(ds^T, da^T, db^T)$ such that for each $t$: $Q(d\pi_t, da_t, db^t) = R^\mu(d\pi_t, da_t, db^t)$.*

**Proof:** For each $t$ choose a channel input distribution that satisfies:

$$q(da_t \mid \pi_t, b^{t-1}) = R^\mu(da_t \mid \pi_t, b^{t-1})$$

for $R^\mu$ almost all $\pi_t, b^{t-1}$. We proceed by induction. For $t = 1$ we have $Q(d\pi_1, da_1, db_1) = r(db_1 \mid \pi_1, a_1) \otimes q(da_1 \mid \pi_1) \otimes \delta_{\{p(ds_1)\}}(d\pi_1) = R^\mu(d\pi_1, da_1, db_1)$. For $t + 1$ we have for any

Borel measurable $\Omega \subset \mathcal{P}(\mathcal{S})$ and all $a_{t+1}, b^{t+1}$:

$$Q(\Omega, a_{t+1}, b^{t+1})$$
$$= \int_{\mathcal{A}} \int_{\mathcal{P}(\mathcal{S})} \int_{\Omega} r(b_{t+1} \mid \pi_{t+1}, a_{t+1}) \, q(a_{t+1} \mid \pi_{t+1}, b^t) \, \delta_{\{\Phi_{\Pi}(\pi_t, a_t, b_t)\}}(d\pi_{t+1}) \, Q(d\pi_t, da_t, b^t)$$
$$\stackrel{(a)}{=} \int_{\mathcal{A}} \int_{\mathcal{P}(\mathcal{S})} \int_{\Omega} r(b_{t+1} \mid \pi_{t+1}, a_{t+1}) \, q(a_{t+1} \mid \pi_{t+1}, b^t) \, \delta_{\{\Phi_{\Pi}(\pi_t, a_t, b_t)\}}(d\pi_{t+1}) \, R^{\mu}(d\pi_t, da_t, b^t)$$
$$= \int_{\mathcal{A}} \int_{\mathcal{P}(\mathcal{S})} \int_{\Omega} r(b_{t+1} \mid \pi_{t+1}, a_{t+1}) \, q(a_{t+1} \mid \pi_{t+1}, b^t) \, \delta_{\{\Phi_{\Pi}(\pi_t, a_t, b_t)\}}(d\pi_{t+1})$$
$$\times \left( \int_{\mathcal{U}} r(b_t \mid \pi_t, a_t) u_t(d\pi_t, da_t) \, R^{\mu}(du_t, b^{t-1}) \right)$$
$$\stackrel{(b)}{=} \int_{\Omega} r(b_{t+1} \mid \pi_{t+1}, a_{t+1}) \, q(a_{t+1} \mid \pi_{t+1}, b^t)$$
$$\times \left( \int_{\mathcal{U}} \int_{\mathcal{A}} \int_{\mathcal{P}(\mathcal{S})} \delta_{\{\Phi_{\Pi}(\pi_t, a_t, b_t)\}}(d\pi_{t+1}) \, r(d\pi_t, da_t \mid u_t, b_t) \, r(b_t \mid u_t) R^{\mu}(du_t, b^{t-1}) \right)$$
$$\stackrel{(c)}{=} \int_{\Omega} r(b_{t+1} \mid \pi_{t+1}, a_{t+1}) \, q(a_{t+1} \mid \pi_{t+1}, b^t) \int_{\mathcal{U}} \gamma_{t+1}[u_t, b_t](d\pi_{t+1}) R^{\mu}(du_t, b^t)$$
$$\stackrel{(d)}{=} \int_{\Omega} r(b_{t+1} \mid \pi_{t+1}, a_{t+1}) \, q(a_{t+1} \mid \pi_{t+1}, b^t) \, R^{\mu}(d\pi_{t+1}, b^t)$$
$$\stackrel{(e)}{=} R^{\mu}(\Omega, a_{t+1}, b^{t+1})$$

where (a) follows from the induction hypothesis, (b) follows from lemma 7.1, (c) follows from equation (25), (d) follows from lemma 7.2, and (e) follows from the choice of channel input distribution. □

**Lemma 7.5** *For every policy $\{\mu_t\}$ satisfying (28) with resulting joint measure $R^{\mu}$ there exists a deterministic policy $\{\bar{\mu}_t\}$ satisfying (28) with resulting joint measure $R^{\bar{\mu}}$ such that for each $t$: $E_{R^{\mu}}[\bar{c}(U_t)] \leq E_{R^{\bar{\mu}}}[\bar{c}(U_t)]$.*

**Proof:** Fix $\{\mu_t\}$. By lemma 7.4 we know there is a channel input distribution $\{q(da_t \mid \pi_t, b^{t-1})\}$ such that for each $t$: $Q(d\pi_t, da_t, db^t) = R^{\mu}(d\pi_t, da_t, db^t)$. By lemma 7.3 we know there is a deterministic policy $\{\bar{\mu}_t\}$ such that for each $t$: $R^{\bar{\mu}}(d\pi_t, da_t, db^t) = Q(d\pi_t, da_t, db^t)$. Hence, for this $\{\bar{\mu}_t\}$, we have $R^{\bar{\mu}}(d\pi_t, da_t, db^t) = R^{\mu}(d\pi_t, da_t, db^t)$.

First note that for each $t$, any Borel measurable $\Omega \in \mathcal{P}(\mathcal{S})$, and all $a_t, b^{t-1}$:

$$\bar{\mu}[b^{t-1}](\Omega, a_t) \, R^{\mu}(b^{t-1}) = \bar{\mu}[b^{t-1}](\Omega, a_t) \, R^{\bar{\mu}}(b^{t-1})$$
$$= R^{\bar{\mu}}(\Omega, a_t, b^{t-1})$$
$$= R^{\mu}(\Omega, a_t, b^{t-1})$$
$$= \int_{\mathcal{U}} u_t(\Omega, a_t) \, R^{\mu}(du_t, b^{t-1})$$

This implies $\bar{\mu}[b^{t-1}](d\pi_t, da_t) = \int_{\mathcal{U}} u_t(d\pi_t, da_t) R^{\mu}(du_t \mid b^{t-1})$ for $R^{\mu}$ almost all $b^{t-1}$.

Now for each $t$:

$$E_{R^\mu}[\bar{c}(U_t)]$$
$$= \int_{\mathcal{U}\times\mathcal{B}^{t-1}} R^\mu(du_t, db^{t-1}) \int_{\mathcal{B}}\int_{\mathcal{A}}\int_{\mathcal{P}(\mathcal{S})} r(db_t \mid \pi_t, a_t) u_t(d\pi_t, da_t) \log \frac{r(b_t \mid \pi_t, a_t)}{\int_{\mathcal{A}}\int_{\mathcal{P}(\mathcal{S})} r(b_t \mid \tilde{\pi}_t, \tilde{a}_t) u_t(d\tilde{\pi}_t, d\tilde{a}_t)}$$
$$\overset{(a)}{\leq} \int_{\mathcal{B}^{t-1}} R^\mu(db^{t-1}) \int\int\int r(b_t \mid \pi_t, a_t) \left( \int_{\mathcal{U}} u_t(d\pi_t, da_t) R^\mu(du_t \mid b^{t-1}) \right)$$
$$\times \log \frac{r(b_t \mid \pi_t, a_t)}{\int\int r(b_t \mid \tilde{\pi}_t, \tilde{a}_t) \left( \int_{\mathcal{U}} u(d\tilde{\pi}_t, d\tilde{a}_t) R^\mu(du_t \mid b^{t-1}) \right)}$$
$$\overset{(b)}{=} \int_{\mathcal{B}^{t-1}} R^\mu(db^{t-1}) \int\int\int r(b_t \mid \pi_t, a_t) \bar{\mu}[b^{t-1}](d\pi_t, da_t)$$
$$\times \log \frac{r(b_t \mid \pi_t, a_t)}{\int\int r(b_t \mid \tilde{\pi}_t, \tilde{a}_t) \bar{\mu}_t[b^{t-1}](d\tilde{\pi}_t, d\tilde{a}_t)}$$
$$\overset{(c)}{=} \int R^{\bar{\mu}}(du_t, db^{t-1}) \int\int\int r(b_t \mid \pi_t, a_t) u_t(d\pi_t, da_t) \log \frac{r(b_t \mid \pi_t, a_t)}{\int\int r(b_t \mid \tilde{\pi}_t, \tilde{a}_t) u_t(d\tilde{\pi}_t, d\tilde{a}_t)}$$
$$= E_{R^{\bar{\mu}}}[\bar{c}(U_t)]$$

where (a) follows from the conditional Jensen's inequality; (b) follows from above; and (c) follows because $R^\mu(db^{t-1}) = R^{\bar{\mu}}(db^{t-1})$ and $\bar{\mu}_t$ is a deterministic policy. $\square$

Thus without loss of generality the policies in the POMDP described in equation (31) can be restricted to be deterministic policies.

**Theorem 7.1** *The two optimization problems given by (18) and (31) have the same optimal cost.*

**Proof:** First note that for any deterministic policy $\{\mu_t\}$ satisfying (28) with resulting joint measure $R^\mu$ and, as given in lemma 7.4, an associated channel input distribution $\{q(da_t \mid \pi_t, b^{t-1})\}$ with associated joint measure $Q$ the following holds for each t:

$$E_{R^\mu}[\bar{c}(U_t)]$$
$$= \int_{\mathcal{U}\times\mathcal{B}^{t-1}} R^\mu(du_t, db^{t-1}) \int\int\int r(db_t \mid \pi_t, a_t) u_t(d\pi_t, da_t) \log \frac{r(b_t \mid \pi_t, a_t)}{\int\int r(b_t \mid \tilde{\pi}_t, \tilde{a}_t) u_t(d\tilde{\pi}_t, d\tilde{a}_t)}$$
$$\overset{(a)}{=} \int_{\mathcal{B}^{t-1}} R^\mu(db^{t-1}) \int\int\int r(db_t \mid \pi_t, a_t) \mu_t[p(ds_1), b^{t-1}](d\pi_t, da_t)$$
$$\times \log \frac{r(b_t \mid \pi_t, a_t)}{\int\int r(b_t \mid \tilde{\pi}_t, \tilde{a}_t) \mu_t[b^{t-1}](d\tilde{\pi}_t, d\tilde{a}_t)}$$
$$\overset{(b)}{=} \int_{\mathcal{B}^{t-1}} R^\mu(db^{t-1}) \int\int\int r(db_t \mid \pi_t, a_t) R^\mu(d\pi_t, da_t \mid b^{t-1}) \log \frac{r(b_t \mid \pi_t, a_t)}{\int\int r(b_t \mid \tilde{\pi}_t, \tilde{a}_t) R^\mu(d\tilde{\pi}_t, d\tilde{a}_t \mid b^{t-1})}$$
$$\overset{(c)}{=} \int_{\mathcal{B}^{t-1}} Q(db^{t-1}) \int\int\int r(db_t \mid \pi_t, a_t) Q(d\pi_t, da_t \mid b^{t-1}) \log \frac{r(b_t \mid \pi_t, a_t)}{\int\int r(b_t \mid \tilde{\pi}_t, \tilde{a}_t) Q(d\tilde{\pi}_t, d\tilde{a}_t \mid b^{t-1})}$$
$$= I_Q(A_t, \Pi_t; B_t \mid B^{t-1})$$
$$\overset{(d)}{=} I_{R^\mu}(A_t, \Pi_t; B_t \mid B^{t-1})$$

where (a) and (b) follow because $\mu_t$ is a deterministic policy and hence: $R^\mu(d\pi_t, da_t \mid b^{t-1})$ $= \mu_t[b^{t-1}](d\pi_t, da_t)$. Lines (c) and (d) follow because $Q(d\pi_t, da_t, db^t) = R^\mu(d\pi_t, da_t, db^t)$. The theorem then follows from this observation and lemmas 7.3-7.5. □

## 7.4 Fully Observed Markov Decision Problem

In this section we make one final simplification. We will convert the POMDP into a fully observed MDP on a suitably defined state space.

Note that the cost, $\bar{c}(u)$ given in equation (30), at time $t$ only depends on $u_t$. The control constraints, $U(\gamma)$ given in equation (26), at time $t$ only depends on $\gamma_t$. The statistics $\gamma_t[u^{t-1}, b^{t-1}]$ only depend on $p(ds_1)$ in the case $t = 1$ and only depends on $u_{t-1}, b_{t-1}$ in the case $t > 1$.

This suggests that $\Gamma_t \in \mathcal{P}(\mathcal{P}(\mathcal{S}))$ could be a suitable fully observed state. The dynamics are given as: $\gamma_1(d\pi_1) = \delta_{\{p(ds_1)\}}(d\pi_1)$ and for $t > 1$:

$$r(d\gamma_{t+1} \mid \gamma_t, u_t) = \int_{\mathcal{P}(\mathcal{S})} \int_\mathcal{A} \int_\mathcal{B} \delta_{\{\Phi_\Gamma(u_t, b_t)\}}(d\gamma_{t+1}) \, r(db_t \mid \pi_t, a_t) \, u_t(d\pi_t, da_t) \quad (33)$$

**Lemma 7.6** *For every policy $\{\mu_t\}$ satisfying (28) with resulting joint measure $R^\mu$ we have for each $t > 1$:*

$$R^\mu(d\gamma_t \mid \gamma_{t-1}, u_{t-1}) = r(d\gamma_t \mid \gamma_{t-1}, u_{t-1}) \quad (34)$$

*for $R^\mu$ almost all $\gamma_{t-1}, u_{t-1}$.*

**Proof:** For each $t > 1$ and for any Borel measurable sets $\Omega_{t-1}, \Omega_t \subset \mathcal{P}(\mathcal{P}(\mathcal{S}))$ and any Borel measurable set $\Theta \subset \mathcal{U}$ we have:

$$R^\mu(\Omega_t, \Omega_{t-1}, \Theta)$$
$$= \int_\Theta \int_{\Omega_{t-1}} \int_\mathcal{B} \int_\mathcal{A} \int_{\mathcal{P}(\mathcal{S})} R^\mu(\Omega_t, d\gamma_{t-1}, du_{t-1}, d\pi_{t-1}, da_{t-1}, db_{t-1})$$
$$= \int \int \int \int \int \{\Phi_\Gamma(u_{t-1}, b_{t-1}) \in \Omega_t\} r(db_{t-1} \mid \pi_{t-1}, a_{t-1}) u_{t-1}(d\pi_{t-1}, da_{t-1}) R^\mu(d\gamma_{t-1}, du_{t-1})$$
$$= \int_\Theta \int_{\Omega_{t-1}} r(\Omega_t \mid \gamma_{t-1}, u_{t-1}) R^\mu(d\gamma_{t-1}, du_{t-1})$$

where the last line follows from equation (33). □

Note that the dynamics given in equation (33), $r(d\gamma_{t+1} \mid \gamma_t, u_t)$ depend only on $u_t$. This along with the fact that the cost at time $t$ only depends on $u_t$ and the control constraint at time $t$ only depends on $\gamma_t$ suggests that we can simplify the form of the control policy from $\mu_t(du_t \mid u^{t-1}, b^{t-1})$ to $\mu_t(du_t \mid \gamma_t)$.

**Theorem 7.2** *Without loss of generality, the optimization given in equation (31) can be modelled as a fully observed MDP with*

(1) *State space $\mathcal{P}(\mathcal{P}(\mathcal{S}))$ and dynamics given by (33)*

(2) *Compact control constraints $U(\gamma)$ given by (26)*

(3) Running cost $\bar{c}(u)$ given by (30).

**Proof:** See section 10.2 of Bertsekas and Schreve [4]. In particular proposition 10.5. □

Theorem 7.1 shows that for any deterministic policy $\{\mu_t[b^{t-1}]\}$ with resulting joint measure $R^\mu$ there is a corresponding channel input distribution $q(da_t \mid \pi_t, b^{t-1})$ with resulting joint measure $Q$ such that for all $t$: $Q(d\pi_t, da_t, db^t) = R^\mu(d\pi_t, da_t, db^t)$. Hence $Q(da_t \mid \pi_t, b^{t-1}) = R^\mu(da_t \mid \pi_t, b^{t-1})$ for $R^\mu$ almost surely all $\pi_t, b^{t-1}$. Theorem 7.1 also shows $\bar{c}(\mu_t[b^{t-1}]) = I_Q(A_t, \Pi_t; B_t \mid b^{t-1})$ $R^\mu$–almost all $b^{t-1}$.

By theorem 7.2. we know we can, without loss of generality, restrict ourselves to deterministic policies of the form: $\{\mu_t[\gamma_t]\}$. Under such a policy we have:

$$Q(da_t \mid \pi_t, b^{t-1}) = R^\mu(da_t \mid \pi_t, b^{t-1}) = R^\mu(da_t \mid \pi_t, \gamma_t)$$

for $R^\mu$ almost surely all $\pi_t, b^{t-1}, \gamma_t$. For a fixed deterministic policy $\gamma_t$ is a function of $b^{t-1}$. Thus the optimal channel input distribution takes the form $\{q(da_t \mid \pi_t, \gamma_t)\}$ and

$$\bar{c}(\mu_t[\gamma_t]) = I_Q(A_t, \Pi_t; B_t \mid \gamma_t) \quad R^\mu - \text{almost all } \gamma_t \tag{35}$$

Recall that in equation (18) we started with terms of the form $I(A^t; B_t \mid B^{t-1})$ and have now simplified it to terms of the form $I(A_t, \Pi_t; B_t \mid \Gamma_t)$.

## 7.5 ACOE and Information Stability

In this section we present the ACOE for the fully observed MDP corresponding to the equivalent optimizations in (18) and (31). We then show that the process is information stable under the optimal input distribution. Finally we relate the equivalent optimizations in (18) and (31) to the optimization given in (4): $\sup_{\{\mathcal{D}_T\}_{T=1}^\infty} \underline{I}(A \to B)$.

The following technical lemma is required to insure the existence of a *measurable selector* in the ACOE given in (36) below. The proof is straightforward but tedious and can be found in the appendix.

**Lemma 7.7** *For $|\mathcal{B}|$ finite we have*

(1) *The cost is bounded and continuous. Specifically, $0 \leq \bar{c}(u) \leq \log |\mathcal{B}|$, $\forall u \in \mathcal{U}$.*

(2) *The control constraint function $\mathcal{U}(\gamma)$ is a continuous set-valued map between $\mathcal{P}(\mathcal{P}(\mathcal{S}))$ and $\mathcal{U}$.*

(3) *The dynamics $r(d\gamma_{t+1} \mid \gamma_t, u_t)$ are continuous.*

We now present the infinite horizon average cost verification theorem.

**Theorem 7.3** *If there exists a $V^* \in \mathbb{R}$, a bounded function $w : \gamma \mapsto w(\gamma) \in \mathbb{R}$, and a policy $\mu^*$ achieving the supremum for each $\gamma$ in the following average cost optimality equation (ACOE):*

$$V^* + w(\gamma) = \sup_{u \in \mathcal{U}(\gamma)} \left( \bar{c}(u) + \int w(\tilde{\gamma}) r(d\tilde{\gamma} \mid \gamma, u) \right) \tag{36}$$

*then*

(1) $V^*$ is the optimal value of the optimization in (31). The optimal policy is the stationary, deterministic policy given by $\mu^*$.

(2) Under this $\mu^*$ we have

$$V^* = \liminf_{T \to \infty} \frac{1}{T} E_{R^{\mu^*}} \left[ \sum_{t=1}^{T} \bar{c}(U_t) \right] = \limsup_{T \to \infty} \frac{1}{T} E_{R^{\mu^*}} \left[ \sum_{t=1}^{T} \bar{c}(U_t) \right]$$

and

$$\lim_{T \to \infty} \frac{1}{T} \sum_{t=1}^{T} \bar{c}(U_t) = V^* \quad R^{\mu^*} - a.s.$$

**Proof:** Follows from lemma 7.7 and theorems 6.2 and 6.3 of [1]. □

For a measure $Q(d\pi_t, da_t, db^t)$ define

$$i_Q(a_t, \pi_t; b_t \mid b^{t-1}) = \log \frac{r(b_t \mid \pi_t, a_t)}{\int r(b_t \mid \tilde{\pi}_t, \tilde{a}_t) Q(d\tilde{\pi}_t, d\tilde{a}_t \mid b^{t-1})} \quad Q - \text{almost all } a_t, \pi_t, b^t \quad (37)$$

The following theorem will allow us to view the ACOE, equation (36), as an implicit single-letter characterization of the capacity of the Markov channel.

**Theorem 7.4** *Assume there exists a $V^* \in \mathbb{R}$, a bounded function $w : \gamma \mapsto w(\gamma) \in \mathbb{R}$, and a policy $\mu^*$ achieving the supremum for each $\gamma$ in ACOE (36). For $\mu^*$ and resulting joint measure $R^{\mu^*}$ let $\{q^*(da_t \mid \pi_t, b^{t-1})\}$ be the corresponding optimal channel input distribution and $Q^*$ be the corresponding measure.*

(1) $\lim_{T \to \infty} \frac{1}{T} \sum_{t=1}^{t} i_{Q^*}(A_t, \Pi_t; B_t \mid B^{t-1}) = V^* \quad Q^* - a.s.$

(2) The channel is directed information stable and has a strong converse under the optimal channel input distribution $\{q^*(da_t \mid \pi_t, b^{t-1})\}$.

(3) $V^* = C = \sup_{\{\mathcal{D}_T\}_{T=1}^{\infty}} \underline{I}(A \to B)$ is the capacity of the channel.

**Proof:** We first prove part (2) and (3) assuming part (1) is true. Part (2) follows from part (1) and proposition 5.1. To prove part (3) note:

$$\begin{aligned}
C &= \sup_{\{\mathcal{D}_T\}_{T=1}^{\infty}} \underline{I}_Q(A \to B) \\
&\overset{(a)}{\leq} \sup_{\{\mathcal{D}_T\}_{T=1}^{\infty}} \liminf_{T \to \infty} \frac{1}{T} I_Q(A^T \to B^T) \\
&\overset{(b)}{=} \sup_{\{\{\mu_t\}_{t=1}^{T}\}_{T=1}^{\infty}} \liminf_{T \to \infty} \frac{1}{T} \sum_{t=1}^{T} E_R \bar{c}(U_t) \\
&\overset{(c)}{=} \sup_{\{\mu_t\}_{t=1}^{\infty}} \liminf_{T \to \infty} \frac{1}{T} \sum_{t=1}^{T} E_R \bar{c}(U_t) \\
&= V^*
\end{aligned}$$

where (a) follows from lemma 4.1; (b) follows from theorems 7.1 and 7.2; and (c) follows from Bellman's principle of optimality. Note the supremizations in (b) and (c) are over policies that satisfy the control constraint (28). Now by part (1) we see that (a) holds with equality. Hence part (3) follows.

We need only prove part (1). Note that theorem 7.3(2) implies:

$$
\begin{aligned}
V^* &= \lim_{T\to\infty} \frac{1}{T} \sum_{t=1}^{T} \bar{c}(U_t) \quad R^{\mu^*} - a.s. \\
&\stackrel{(a)}{=} \lim_{T\to\infty} \frac{1}{T} \sum_{t=1}^{T} I_{Q^*}(A_t, \Pi_t; B_t \mid b^{t-1}) \quad R^{\mu^*} - \text{almost all } b^\infty \\
&\stackrel{(b)}{=} \lim_{T\to\infty} \frac{1}{T} \sum_{t=1}^{T} I_{Q^*}(A_t, \Pi_t; B_t \mid b^{t-1}) \quad Q^* - \text{almost all } b^\infty
\end{aligned}
$$

where (a) follows from (35) and (b) follows because for each $t$: $Q^*(db^t) = R^{\mu^*}(db^t)$. Hence $Q^*(db^\infty) = R^{\mu^*}(db^\infty)$.

Define the nested family of sigma-fields: $\mathbb{F}_t = \sigma(\Pi^t, A^t, B^t)$. Let

$$Z_t(\pi_t, a_t, b^t) = i_{Q^*}(a_t, \pi_t; b_t \mid b^{t-1}) - I_{Q^*}(A_t, \Pi_t; B_t \mid b^{t-1})$$

Clearly $Z_t$ is $\mathbb{F}_t$-measurable and $E_{Q^*}(Z_t \mid \mathbb{F}_{t-1}) = 0$  $Q^* - a.s.$ Hence $Z_t$ is a martingale difference sequence. The martingale stability theorem states if

$$\lim_{T\to\infty} \sum_{t=1}^{T} \frac{E_{Q^*}[Z_t^2 \mid \mathbb{F}_{t-1}]}{t^2} < \infty \quad Q^* - a.s. \tag{38}$$

then $\lim_{T\to\infty} \frac{1}{T} \sum_{t=1}^{t} Z_t = 0$  $Q^* - a.s.$ This in turn would imply

$$\lim_{T\to\infty} \frac{1}{T} \sum_{t=1}^{t} i_{Q^*}(a_t, \pi_t; b_t \mid b^{t-1}) = \lim_{T\to\infty} \frac{1}{T} \sum_{t=1}^{t} I_{Q^*}(A_t, \Pi_t; B_t \mid b^{t-1}) = V^*$$

for $Q^* - $ almost all $\pi^\infty, a^\infty, b^\infty$.

To show that (38) holds note that for any $t$ and $Q^*$-almost all $b^{t-1}$ we have:

$$
\begin{aligned}
&E_{Q^*}(Z_t^2 \mid b^{t-1}) \\
\leq\; & E_{Q^*}[i_{Q^*}^2(A_t, \Pi_t; B_t \mid B^{t-1}) \mid b^{t-1}] \\
=\; & E_{Q^*}\left[ \log^2 r(B_t \mid \Pi_t, A_t) + \log^2\left( \int r(B_t \mid \tilde{\pi}_t, \tilde{a}_t) Q^*(d\tilde{\pi}_t, d\tilde{a}_t \mid B^{t-1}) \right) \right. \\
& \left. - 2 \log r(B_t \mid \Pi_t, A_t) \log\left( \int r(B_t \mid \tilde{\pi}_t, \tilde{a}_t) Q^*(d\tilde{\pi}_t, d\tilde{a}_t \mid B^{t-1}) \right) \mid b^{t-1} \right] \\
\leq\; & E_{Q^*}\left[ \log^2 r(B_t \mid \Pi_t, A_t) \mid b^{t-1} \right] + E_{Q^*}\left[ \log^2\left( \int r(B_t \mid \tilde{\pi}_t, \tilde{a}_t) Q^*(d\tilde{\pi}_t, d\tilde{a}_t \mid B^{t-1}) \right) \mid b^{t-1} \right] \\
\leq\; & 2|\mathcal{B}|
\end{aligned}
$$

The last inequality follows because the function $x \log^2 x$ achieves a maximum value of 1 over the domain $0 \leq x \leq 1$. Since $\sum_t \frac{2|\mathcal{B}|}{t^2}$ is summable we see that (38) holds. $\square$

There remains the question of when a solution to the ACOE exists. There exist many sufficient conditions for the existence of a solution. See [1], [17] for a representative sample. Most of these conditions require the process be recurrent under the optimal policy. The following theorem describes one such sufficient condition:

**Theorem 7.5** *If there exists an $\alpha < 1$ such that*

$$\sup_{\gamma_t, \tilde{\gamma}_t, u_t \in \mathcal{U}(\gamma_t), \tilde{u}_t \in \mathcal{U}(\tilde{\gamma}_t)} \|r(d\gamma_{t+1} \mid \gamma_t, u_t) - r(d\gamma_{t+1} \mid \tilde{\gamma}_t, \tilde{u}_t)\|_{TV} \leq \alpha \quad (39)$$

*then the ACOE (36) has a bounded solution. Here $\|\cdot\|_{TV}$ denotes the total variation norm.*

**Proof:** See corollary 6.1 of [1]. □

Condition (39) insures that for any stationary policy there exists a stationary distribution. Specifically:

**Proposition 7.1** *If (39) holds then for all stationary policies of the form, $\mu : \gamma \to u(d\pi, da)$, there exists a probably measure $\lambda_\mu$ on $\mathcal{P}(\mathcal{S})$ such that for any $\epsilon > 0$ there exists a $T$ large enough such that $\forall t > T$:*

$$\|r_\mu^t(d\gamma_t \mid \gamma_1) - \lambda_\mu(d\gamma_t)\|_{TV} \leq \epsilon \quad (40)$$

*where $r_\mu^t(d\gamma_t \mid \gamma_1)$ is the $t-$step transition stochastic kernel under the stationary policy $\mu$. Furthermore $\lim_{T\to\infty} \frac{1}{T} E_{R^\mu} \left[\sum_{t=1}^T \bar{c}(\mu(\Gamma_t))\right] = \int \bar{c}(\mu(\gamma)) \lambda_\mu(d\gamma)$ independent of the choice of $p(ds_1)$.*

**Proof:** See lemma 3.3 of [17]. □

We have until this point assumed that the Markov channel parameters are fixed. The last part of proposition 7.1 shows that the capacity $C$ is the same no matter which choice of $p(ds_1)$ is chosen.

In the case that one chooses a policy without feedback equation (40) essentially reduces to the definition of indecomposability found in Gallager [14], equation 4.6.26.

Finding conditions that imply (39) or (40) directly in terms of the Markov channel, $\{p(ds_1), p(ds_{t+1} \mid s_t, a_t, b_t), p(db_t \mid s_t, a_t)\}$, is challenging. This is essentially the problem of determining conditions for the ergodicity of the underlying hidden Markov model under the optimal stationary policy. See [18], [2], [21], [10] for some representative conditions.

## 8  Cases with Simple Sufficient Statistics

As we have already seen the sufficient statistics $\Pi_t \in \mathcal{P}(\mathcal{S})$ and $\Gamma_t \in \mathcal{P}(\mathcal{P}(\mathcal{S}))$ can be quite complicated in general. There are, though, many situations where they become much simpler.

### 8.1  $S$ Computable from the Channel Input and Output

Recall that $\Pi_t$ is a function of $(A^{t-1}, B^{t-1})$ and satisfies the recursion: $\pi_{t+1} = \Phi_\Pi(\pi_t, a_t, b_t)$. In many scenarios the state $S_t$ is computable from $(A^{t-1}, B^{t-1})$. Here we assume that $p(ds_1) = \delta_{\{s_1\}}(ds_1)$ for some fixed state $s_1$ and for $t > 1$: $\Pi_t(ds_t) = \delta_{\{S_t\}}(ds_t)$ $Q-a.s.$

This in turn implies there exists a function $\Phi_S$ such that $s_{t+1} = \Phi_S(s_t, a_t, b_t)$. To see this recall equation (13). Because $\Pi_t, \Pi_{t+1}$ are Diracs $Q$-almost surely it must be the case that $S_{t+1}$ is a function of $S_t, A_t, B_t$ $Q - a.s.$ One example of such a channel would be: $\{p(db_t \mid a_t, a_{t-1}, b_{t-1})\}$. Here one could choose the state: $S_t = (A_{t-1}, B_{t-1})$.

We can directly associate $\Pi$ with $S$. In addition $\Gamma$ can be viewed as a conditional probability of the state $S$. Hence we can restrict ourselves to control policies of the form: $\mu : \mathcal{P}(\mathcal{S}) \to \mathcal{U} = \mathcal{P}(\mathcal{S} \times \mathcal{A})$ taking $\gamma \mapsto u(ds, da)$. Now the control constraints take the form $\mathcal{U}(\gamma) = \{u(ds, da) : u(ds, da) \in \mathcal{U}, u(ds) = \gamma(ds)\}$. The dynamics of $\Gamma_t$ given in equations (24), (25) simplify to: $\gamma_1(ds_1) = \delta_{\{s_1\}}(ds_1)$ and for $t > 1$ and all $s_t$:

$$\gamma_t[u^{t-1}, b^{t-1}](s_t) = \sum_{s_{t-1}, a_{t-1}} \delta_{\{\Phi_S(s_{t-1}, a_{t-1}, b_{t-1})\}}(s_t) \, r(ds_{t-1}, a_{t-1} \mid u_{t-1}, b_{t-1}) \quad (41)$$

Hence equation (33) simplifies to:

$$r(d\gamma_{t+1} \mid \gamma_t, u_t) = \sum_{s_t, a_t, b_t} \delta_{\{\Phi_\Gamma(u_t, b_t)\}}(d\gamma_{t+1}) \, p(b_t \mid s_t, a_t) \, u_t(s_t, a_t) \quad (42)$$

where $\Phi_\Gamma(u, b)$ comes from (41). The cost in equation (30) simplifies as well:

$$\bar{c}(u) = \sum_{s,a,b} p(b \mid s, a) u(s, a) \log \frac{p(b \mid s, a)}{\sum_{\tilde{s}, \tilde{a}} p(b \mid \tilde{s}, \tilde{a}) u(\tilde{s}, \tilde{a})} \quad (43)$$

In addition
$$I(A_t, \Pi_t; B_t \mid \Gamma_t) = I(A_t; B_t \mid S_t, \Gamma_t) + I(S_t; B_t \mid \Gamma_t) \quad (44)$$

Finally the ACOE equation (36) in theorem 7.3 simplifies to an equation where $w(\gamma)$ is now a function over $\mathcal{P}(\mathcal{S})$:

$$V^* + w(\gamma) = \sup_{u \in \mathcal{U}(\gamma)} \left( \bar{c}(u) + \int w(\tilde{\gamma}) r(d\tilde{\gamma} \mid \gamma, u) \right) \quad (45)$$

The sufficient condition, equation (39), given in theorem 7.5 continues to hold with dynamics given by (42).

We now examine two cases where the computations simplify further: $S$ is either computable from the channel output or channel input only.

### 8.1.1 Case 1: $S$ Computable from the Channel Input Only

Here we assume $S_t$ is computable from only $A^{t-1}$ and hence $S$ is known to the transmitter. Hence $\Pi_t$ is a function of $A^{t-1}$ and satisfies the recursion: $\pi_{t+1} = \Phi_\Pi(\pi_t, a_t)$. This in turn implies there exists a function $\Phi_S$ such that $s_{t+1} = \Phi_S(s_t, a_t)$. These channels are often called *finite state machine Markov channels*. Note that any general channel of the form $\{p(db_t \mid a_t, a_{t-\Delta}^{t-1})\}$, for a finite $\Delta$, can be converted into a Markov channel with state, $S_t = A_{t-\Delta}^{t-1}$, computable from the channel input.

As before we can directly associate $\Pi$ with $S$ and $\Gamma$ can be viewed as a conditional probability of the state $S$. Equations (41)-(45) continue to hold with obvious modifications. See [38], [39] for more details. For Gaussian finite state machine Markov channels the estimate $\Gamma_t$ can be easily computed by using a Kalman filter [40], [41].

### 8.1.2 Case 2: $S$ Computable from the Channel Output Only

Here we assume $S_t$ is computable from only $B^{t-1}$. Thus $S$ is known to the receiver, and via feedback, is known to the transmitter. Then $\Pi_t$ is a function of $B^{t-1}$ and satisfies the recursion: $\pi_{t+1} = \Phi_\Pi(\pi_t, b_t)$. This in turn implies there exists a function $\Phi_S$ such that $s_{t+1} = \Phi_S(s_t, b_t)$. Note that any general channel of the form $\{p(db_t \mid a_t, b_{t-\Delta}^{t-1})\}$, for a finite $\Delta$, can be converted into a Markov channel with state, $S_t = B_{t-\Delta}^{t-1}$, computable from the channel output.

As before we can directly associate $\Pi$ with $S$. In addition, because $\Pi$ is computable from the channel outputs we can directly associate $\Gamma$ with $\Pi$ and hence with $S$. We can then restrict ourselves to control policies of the form: $\mu : \mathcal{S} \to \mathcal{U} = \mathcal{P}(\mathcal{A})$ taking $\gamma \mapsto u(da)$. To see this note that the control constraints become trivial and hence we can use control actions of the form $u(da)$ as opposed to $u(ds, da)$. In this case the dynamics in (33) simplify quite a bit: $\gamma_1 = s_1$ and for $t > 1$:

$$r(d\gamma_{t+1} \mid \gamma_t, u_t) = \sum_{a_t, b_t} \delta_{\{\Phi_S(\gamma_t, b_t)\}}(d\gamma_{t+1}) \, p(b_t \mid \gamma_t, a_t) \, u_t(a_t) \tag{46}$$

The cost in equation (30) simplifies as well:

$$\bar{c}(s, u) = \sum_{a,b} p(b \mid s, a) u(a) \log \frac{p(b \mid s, a)}{\sum_{\tilde{a}} p(b \mid s, \tilde{a}) u(\tilde{a})}$$

In addition $I(A_t, \Pi_t; B_t \mid \Gamma_t) = I(A_t, S_t; B_t \mid S_t) = I(A_t; B_t \mid S_t)$. Finally the ACOE equation (36) in theorem 7.3 simplifies to an equation where $w(\gamma)$ is now a function over $\mathcal{S}$:

$$V^* + w(\gamma) = \sup_{u \in \mathcal{U}} \left( \bar{c}(\gamma, u) + \int w(\tilde{\gamma}) r(d\tilde{\gamma} \mid \gamma, u) \right) \tag{47}$$

**Markov Channels with State Observable to the Receiver:** An important scenario that falls under the case just described is that of a Markov channel, $p(ds_1)$, $\{p(ds_{t+1} \mid s_t, a_t, b_t)\}$, $\{p(db_t \mid s_t, a_t)\}$ with state observable to the receiver. Specifically at time $t$ we assume that along with $B_t$, the state $S_{t+1}$ is observable to the receiver. The standard technique for dealing with this setting is to define a new channel output as follows: $\bar{B}_t = (B_t, S_{t+1})$. The new Markov channel has the same state transition kernel but the channel output is: $p(d\bar{b}_t \mid s_t, a_t) = p(ds_{t+1} \mid s_t, a_t, b_t) \otimes p(db_t \mid s_t, a_t)$. We also assume that $S_1$ is observable to the transmitter. (This can be achieved by assuming that $\bar{B}_0 = S_1$ is transmitted during epoch 0.) Thus the dynamics in (46) can be written as: $\gamma_1 = s_1$ and for $t > 1$:

$$r(\gamma_{t+1} \mid \gamma_t, u_t) = \sum_{a_t, b_t} p(\gamma_{t+1} \mid \gamma_t, a_t, b_t) \, p(b_t \mid \gamma_t, a_t) \, u_t(a_t) \tag{48}$$

Also, $I(A_t, \Pi_t; \bar{B}_t \mid \Gamma_t) = I(A_t; B_t \mid S_t) + I(A_t; S_{t+1} \mid S_t, B_t)$. The second addend is zero if there is no ISI.

The sufficient condition, equation (39), given in theorem 7.5 continues to hold with dynamics given by (48). If there is no ISI then equation (48) reduces to: $r(d\gamma_{t+1} \mid \gamma_t, u_t) =$ $p(d\gamma_{t+1} \mid \gamma_t)$. If $p(ds_{t+1} \mid s_t)$ is an ergodic transition kernel with stationary distribution $\nu$ then there exists a bounded solution to the ACOE [1]. The ACOE reduces to: $V^* = \sum_s \nu(s) \max_u \bar{c}(s, u)$. Thus we recover the well known formula for the capacity of a non-ISI ergodic Markov channel with state available to both the transmitter and receiver.

## 8.2  Π Computable from the Channel Output

Here we assume that $\Pi_t$ is a function of $B^{t-1}$ only and thus satisfies the recursion: $\pi_{t+1} = \Phi_\Pi(\pi_t, b_t)$. Hence $\Gamma_t(d\pi_t) = \delta_{\{\Pi_t\}}(d\pi_t)$ $Q - a.s.$. We can thus directly associate $\Gamma$ with $\Pi$. Now $\Gamma$ can be viewed as a conditional probability of the state $S$. One can view the associated canonical Markov channel as a Markov channel with state $\Pi$ computable from the channel output only (as discussed in the previous section.)

We can then restrict ourselves to control policies of the form: $\mu : \mathcal{P}(\mathcal{S}) \to \mathcal{U} = \mathcal{P}(\mathcal{A})$ taking $\gamma \mapsto u(da)$. To see this note that the control constraints become trivial and hence we can use control actions of the form $u(da)$ as opposed to $u(d\pi, da)$. In this case the dynamics in (33): $\gamma_1(d\pi_1) = \delta_{\{p(ds_1)\}}(d\pi_1)$ and for $t > 1$:

$$r(d\gamma_{t+1} \mid \gamma_t, u_t) = \sum_{s_t, a_t, b_t} \delta_{\{\Phi_\Pi(\gamma_t, b_t)\}}(d\gamma_{t+1}) \; p(b_t \mid s_t, a_t) \; \gamma_t(s_t) \; u_t(a_t) \tag{49}$$

The cost in equation (30) simplifies as well:

$$\bar{c}(\pi, u) = \sum_{s,a,b} p(b \mid s, a) \; \pi(s) \; u(a) \log \frac{\sum_{\tilde{s}} p(b \mid \tilde{s}, a) \; \pi(\tilde{s})}{\sum_{\tilde{s}, \tilde{a}} p(b \mid \tilde{s}, \tilde{a}) \; \pi(\tilde{s}) \; u(\tilde{a})}$$

In addition $I(A_t, \Pi_t; B_t \mid \Gamma_t) = I(A_t; B_t \mid \Pi_t)$. Finally the ACOE equation (36) in theorem 7.3 simplifies to an equation where $w(\gamma)$ is now a function over $\mathcal{P}(\mathcal{S})$:

$$V^* + w(\gamma) = \sup_{u \in \mathcal{U}} \left( \bar{c}(\gamma, u) + \int w(\tilde{\gamma}) r(d\tilde{\gamma} \mid \gamma, u) \right) \tag{50}$$

In this case the optimal channel input distribution $q(da_t \mid \pi_t, \gamma_t)$ can be written in the form $q(da_t \mid b^{t-1})$. Furthermore the code-function distribution can be taken to be a product distribution. Choose for each $t$ and $f_t$:

$$p(f_t) = \prod_{b^{t-1}} q(f_t(b^{t-1}) \mid b^{t-1}).$$

Then $P_{F^T}(df^T) = \otimes_{t=1}^T p(df_t)$. One can easily verify for each $t$ that $P_{F^T}(\Upsilon^t(b^{t-1}, a^t)) = \vec{q}(a^t \mid b^{t-1})$ and hence is good with respect to $\{q(da_t \mid b^{t-1})\}$.

In summary, if the sufficient statistic $\Pi$ is computable from the channel output then the optimal code-function distribution can be taken to be a product measure. If $\Pi_t$ depends on $A^{t-1}$ then the optimal code-function, in general, will not be a product measure.

In this section we discussed scenarios where the sufficient statistics had special structure. One open question is to determine whether the sufficient statistics will simplify if we restrict ourselves to special classes of code-functions. As an example, it was shown in [15] that for non-ISI Markov channels with no feedback one can find a single-letter formula when the channel inputs are independent and identically distributed.

## 9 Maximum Likelihood Decoding

We now consider the problem of maximum-likelihood decoding. For a given message set $\mathcal{W}$ fix a channel code $\{f^T[w]\ w \in \mathcal{W}\}$. Assume the messages are chosen uniformly. Hence each channel code-function is chosen with probability $P_{F^T}(f^T[w]) = \frac{1}{|\mathcal{W}|}$. For a consistent joint measure $Q(df^T, da^T, db^T)$ our task is to simplify the computation of $\arg\max_{w \in \mathcal{W}} Q(f^T[w] \mid b^T)$.

First consider the general channels described in section 5. Note

$$\begin{aligned} Q(f^T, b^T) &= Q(f^T, a^T = f^T(b^{T-1}), b^T) \\ &= Q(f^T \mid a^T = f^T(b^{T-1}), b^T)\ Q(a^T = f^T(b^{T-1}), b^T) \end{aligned} \quad (40)$$

Also, if $Q(a^T, b^T) > 0$ then

$$\begin{aligned} Q(f^T \mid a^T, b^T) &= \frac{Q(f^T, a^T, b^T)}{Q(a^T, b^T)} \\ &= \frac{P_{F^T}(f^T)\vec{p}(b^T \mid a^T) \prod_{t=1}^T \delta_{\{f_t(b^{t-1})\}}(a_t)}{\vec{p}(b^T \mid a^T) \prod_{t=1}^T Q(a_t \mid a^{t-1}, b^{t-1})} \\ &\stackrel{(a)}{=} \frac{P_{F^T}(f^T)}{P_{F^T}(\Upsilon^T(b^{T-1}, a^T))} \\ &= \frac{1}{|\Upsilon^T(b^{T-1}, a^T)|} \end{aligned} \quad (41)$$

where (a) follows by lemma 5.1. Note this implies that $F^T - A^T - B^T$ is *not* a Markov chain under $Q$.

Due to the feedback we effectively have a different channel code without feedback for each $b^{T-1}$. Specifically, for each $b^{T-1}$ define

$$\Lambda(b^{T-1}) = \{a^T\ :\ a^T = f^T[w](b^{T-1})\text{ for some }w \in \mathcal{W}\}.$$

From equations (40) and (41) we see that computing $\arg\max_{w \in \mathcal{W}} Q(f^T[w] \mid b^T)$ is equivalent to computing:

$$\arg\max_{a^T \in \Lambda(b^{T-1})} \frac{1}{|\Upsilon^T(b^{T-1}, a^T)|} \prod_{t=1}^T p(b_t \mid a^t, b^{t-1})\ Q(a_t \mid a^{t-1}, b^{t-1}) \quad (42)$$

where $\{Q(da_t \mid a^{t-1}, b^{t-1})\}$ is the induced channel input distribution for $P_{F^T}(f^T[w])$.

For the Markov channel case we may replace $p(db_t \mid a^t, b^{t-1})$ with $p(db_t \mid \pi_t, a_t)$ in equation (42). If in addition, the channel code is chosen such that the induced channel input distribution has the form $\{q(da_t \mid \pi_t, \gamma_t)\}$ then:

$$\arg\max_{a^T \in \Lambda(b^{T-1})} \frac{1}{|\Upsilon^T(b^{T-1}, a^T)|} \prod_{t=1}^T p(b_t \mid \pi_t, a_t)\ q(a_t \mid \pi_t, \gamma_t) \quad (43)$$

In the case where $|\Upsilon^T(b^{T-1}, a^T)| = 1$, $\forall a^T \in \Lambda(b^{T-1})$ the optimization in (43) can be treated as a deterministic longest path problem.

# 10 Conclusion

We have presented a general framework for treating channels with memory and feedback. We first proved a general coding theorem based on Massey's concept of *directed information* and Dobrushin's program of communication as interconnection. We then specialized this result to the case of Markov channels. To compute the capacity of these Markov channels we converted the directed information optimization problem into a partially observed MDP. This required identifying appropriate sufficient statistics at the encoder and decoder. The ACOE verification theorem was presented and sufficient conditions for the existence of a solution were provided. The complexity of many feedback problems can now be understood by examining the complexity of the associated ACOE.

The framework developed herein leaves open the possibility of using approximate dynamic programming techniques, like value and policy iteration and reinforcement learning, for computing the capacity. In addition the framework allows one to compute the capacity under restricted policies. This is useful if one is willing to sacrifice capacity for the benefit of a simpler policy.

**Acknowledgements:** The authors would like to thank Vivek Borkar for many helpful discussions.

# A  Appendix

## A.1  Review of Stochastic Kernels

The following results are standard and can be found in, for example, [4]. Let $(\mathcal{V}, \mathcal{A})$ be a Borel space and let $(\mathcal{X}, \mathcal{B}_\mathcal{X})$ and $(\mathcal{Y}, \mathcal{B}_\mathcal{Y})$ be Polish spaces equipped with their Borel $\sigma$-algebras.

**Definition A.1** *Let $\tau(dx \mid v)$ be a family of probability measures on $\mathcal{X}$ parameterized by $v \in \mathcal{V}$. We say that $\tau$ is a stochastic kernel from $\mathcal{V}$ to $\mathcal{X}$ if for every Borel set $B \in \mathcal{B}_\mathcal{X}$, the function $v \mapsto \tau(B \mid v) \in [0,1]$ is measurable.*

**Lemma A.1** *For $B \in \mathcal{B}_\mathcal{X}$, define $f_B : \mathcal{P}(\mathcal{X}) \to [0,1]$ by $f_B : \mu \mapsto \mu(B)$ for $\mu \in \mathcal{P}(P\mathcal{X})$. Then*
$$\mathcal{B}_{\mathcal{P}(\mathcal{X})} = \sigma[\cup_{B \in \mathcal{B}_\mathcal{X}} f_B^{-1}(\mathcal{B}_{I\!R})]$$

**Theorem A.1** *Let $\tau(dx \mid v)$ be a family of probability measures on $\mathcal{X}$ given $\mathcal{V}$. Then $\tau(dx \mid v)$ is a stochastic kernel if and only if $v \in \mathcal{V} \in \mathcal{P}(\mathcal{X})$ is measurable. That is if and only if $\tau(\cdot \mid v)$ is a random variable from $\mathcal{V}$ into $\mathcal{P}(\mathcal{X})$.*

Since $\tau(\cdot \mid v)$ is a random variable from $\mathcal{V}$ into $\mathcal{P}(\mathcal{X})$ it follows that the class of stochastic kernels is closed under weak limits (weak topology on the space of probability measures.)

We now discuss interconnections of stochastic kernels. Let $\tau_1(dx \mid v)$ be a stochastic kernel from $\mathcal{V}$ to $\mathcal{X}$ and $\tau_2(dy \mid v, x)$ be a stochastic kernel from $\mathcal{V} \times \mathcal{X}$ to $\mathcal{Y}$. Then the joint stochastic kernel $\tau_1 \otimes \tau_2$ from $\mathcal{V}$ to $\mathcal{X} \times \mathcal{Y}$ is for all $v \in \mathcal{V}$, $A \in \mathcal{B_X}$, and $B \in \mathcal{B_Y}$ we have

$$\tau_1 \otimes \tau_2(A \times B \mid v) = \int_{A \times B} \tau_2(dy \mid v, x)\tau_1(dx \mid v) = \int_A \tau_2(B \mid v, x)\tau_2(dx \mid v).$$

Via the Ionescu-Tulcea theorem this can be generalized to interconnections of countable number of stochastic kernels.

We now discuss the decompositions of measures.

**Theorem A.2** *Let $\lambda(dx \otimes dy)$ be a probability measure on $(\mathcal{X} \times \mathcal{Y}, \mathcal{B_X} \otimes \mathcal{B_Y})$. Let $\lambda_1(A) = \lambda(A, \mathcal{Y})$, $A \in \mathcal{B_X}$ be the first marginal. Then there exists a stochastic kernel $\lambda(dy \mid x)$ on $\mathcal{Y}$ given $\mathcal{X}$ such that for all $A \in \mathcal{B_X}$ and $B \in \mathcal{B_Y}$ we have*

$$\lambda(A \times B) = \int_{A \times B} \lambda_1(dx)\lambda(dy \mid x) = \int_A \lambda(B \mid x)\lambda_1(dx)$$

This can be generalized to a parametric dependence:

**Theorem A.3** *Let $\lambda(dx \otimes dy \mid v)$ be a stochastic kernel on $\mathcal{X} \times \mathcal{Y}$ given $\mathcal{V}$. Let $\lambda_1(A \mid v)$ be the first marginal which is a stochastic kernel on $\mathcal{X}$ given $\mathcal{V}$ defined by*

$$\lambda_1(A \mid v) = \lambda(A, \mathcal{Y} \mid v), \ A \in \mathcal{B_X}, \ v \in \mathcal{V}.$$

*Then there exists a stochastic kernel $\lambda(dy \mid v, x)$ on $\mathcal{Y}$ given $\mathcal{V} \times \mathcal{X}$ such that $\forall v \in \mathcal{V}$, $A \in \mathcal{B_X}$, and $B \in \mathcal{B_Y}$ we have*

$$\lambda(A \times B \mid v) = \int_{A \times B} \lambda_1(dx \mid v)\lambda(dy \mid v, x) = \int_A \lambda(B \mid v, x)\lambda_1(dx \mid v)$$

Let $\lambda(dx \otimes dy \mid v)$ be a stochastic kernel on $\mathcal{X} \times \mathcal{Y}$ given $\mathcal{V}$ and suppose the stochastic kernel $\tau(dy \mid v, x)$ on $\mathcal{Y}$ given $\mathcal{V} \times \mathcal{X}$ satisfies:

$\forall v \in \mathcal{V}, \forall B \in \mathcal{B_Y}$ we have $\lambda(B \mid v, x) = \tau(B \mid v, x)$ for $\lambda_1(dx \mid v)$ almost all $x$.

Then for any measurable function $g : \mathcal{V} \times \mathcal{X} \times \mathcal{Y} \to \mathbb{R}$ and all $v \in \mathcal{V}$ we have

$$E(g(V, X, Y) \mid v) = \int_{\mathcal{X} \times \mathcal{Y}} g(v, x, y)\tau(dy \mid v, x)\lambda_1(dx \mid v)$$

whenever the conditional expectation on the left hand side exists.

Finally, recall that a stochastic kernel $\tau(dy \mid x)$ on $\mathcal{Y}$ given $\mathcal{X}$ is *continuous* if for all continuous bounded functions $v$ on $\mathcal{Y}$ the function $\int v(y)p(dy \mid x)$ is a continuous and bounded function on $\mathcal{X}$.

**Theorem A.4** *If $\tau(dy \mid x)$ is a continuous stochastic kernel on $\mathcal{Y}$ given $\mathcal{X}$ and $v(x, y)$ is a continuous bounded function on $\mathcal{X} \times \mathcal{Y}$ then $\int v(x, y)\tau(dy \mid x)$ is a continuous bounded function on $\mathcal{X}$.*

## A.2 Lemma 4.1

The proof of lemma 4.1 below is adapted from lemma A1 of [16] and theorem 8 of [33]. We need the following three lemmas. Combined they state that the mass of $\vec{i}(A^T;B^T)$ at the tails is small. Recall that $|\mathcal{B}| < \infty$.

**Lemma A.2** *Let $L > \log|\mathcal{B}|$. For any sequence of measures $\{P_{B^T}\}_{t=1}^T$ we have*

$$\lim_{T \to \infty} E\left[\frac{1}{T} \log \frac{1}{P(B^T)} 1_{\{\frac{1}{T}\log \frac{1}{P(B^T)} \geq L\}}\right] = 0.$$

**Proof:** Let $\Omega = \{b^T : P(b^T) \leq 2^{-TL}\}$. Now

$$\begin{aligned}
E\left[\frac{1}{T}\log\frac{1}{P(B^T)} 1_{\{\frac{1}{T}\log\frac{1}{P(B^T)}\geq L\}}\right] &= \frac{1}{T}\sum_{b^T\in\Omega} P(b^T)\log\frac{1}{P(b^T)} \\
&= \frac{1}{T}P(\Omega)\sum_{b^T\in\Omega}\frac{P(b^T)}{P(\Omega)}\log\frac{1}{\frac{P(b^T)}{P(\Omega)}} - \frac{1}{T}P(\Omega)\log P(\Omega) \\
&\leq \frac{1}{T}P(\Omega)\log|\mathcal{B}^T| - \frac{1}{T}P(\Omega)\log P(\Omega) \\
&\leq \frac{1}{T}P(\Omega)\log|\mathcal{B}^T| + \frac{1}{2T}
\end{aligned}$$

where the first inequality follows because entropy is maximized by the uniform distribution and the second inequality follows because $-x\log x \leq \frac{1}{2}$, $0 \leq x \leq 1$. Now $P(\Omega) \leq |\Omega|2^{-TL} \leq |\mathcal{B}^T|2^{-TL}$. Thus $E\left[\frac{1}{T}\log\frac{1}{P(B^T)} 1_{\{\frac{1}{T}\log\frac{1}{P(B^T)}\geq L\}}\right] \leq \log|\mathcal{B}|2^{-T(L-\log|\mathcal{B}|)} + \frac{1}{2T}$. This upper bound goes to zero as $T \to \infty$. □

**Lemma A.3** *For any sequence of joint measures $\{P_{A^T,B^T}\}_{T=1}^\infty$ we have*

$$\lim_{T\to\infty} E\left[\frac{1}{T}\vec{i}(A^T;B^T) 1_{\{\frac{1}{T}\vec{i}(A^T;B^T) \leq 0\}}\right] = 0$$

**Proof:** Follows from page 10 of Pinsker [25].

**Lemma A.4** *Let $L > \log|\mathcal{A}|$. For any sequence of joint measures $\{P_{A^T,B^T}\}_{T=1}^\infty$ we have*

$$\lim_{T\to\infty} E\left[\frac{1}{T}\vec{i}(A^T;B^T) 1_{\{\frac{1}{T}\vec{i}(A^T;B^T) \geq L\}}\right] = 0.$$

**Proof:** Let $\Omega = \{b^T : P(b^T) \leq 2^{-TL}\}$. Note that $\frac{1}{P(B^T)} \geq \frac{\vec{p}(B^T \mid A^T)}{P(B^T)}$ $P_{A^T,B^T} - a.s.$ Now

$$\begin{aligned}
E\left[\frac{1}{T}\vec{i}(A^T,B^T) 1_{\{\frac{1}{T}\vec{i}(A^T;B^T) \geq L\}}\right] &= E\left[\frac{1}{T}\log\frac{\vec{p}(B^T\mid A^T)}{P(B^T)} 1_{\{\frac{1}{T}\log\frac{\vec{p}(B^T\mid A^T)}{P(B^T)} \geq L\}}\right] \\
&\leq E\left[\frac{1}{T}\log\frac{1}{P(B^T)} 1_{\{\frac{1}{T}\log\frac{1}{P(B^T)}\geq L\}}\right] \\
&\leq \left(\log|\mathcal{B}|2^{-T(L-\log|\mathcal{B}|)} + \frac{1}{2T}\right)
\end{aligned}$$

The last inequality follows from lemma A.2. This upper bound goes to zero as $T \to \infty$. □

**Proof of Lemma 4.2:** The second inequality is obvious. To prove the first inequality note $\forall \epsilon > 0$ we have

$$\frac{1}{T} I(A^T \to B^T) \geq E\left[\frac{1}{T} \log \frac{\vec{p}(B^T \mid A^T)}{P(B^T)} 1_{\left\{\frac{1}{T} \log \frac{\vec{p}(B^T \mid A^T)}{P(B^T)} \leq 0\right\}}\right]$$

$$+ 0 \times P\left[0 \leq \frac{1}{T} \log \frac{\vec{p}(B^T \mid A^T)}{P(B^T)} \leq \underline{I}(A \to B) - \epsilon\right]$$

$$+ (\underline{I}(A \to B))P\left[\frac{1}{T} \log \frac{\vec{p}(B^T \mid A^T)}{P(B^T)} \geq \underline{I}(A \to B) - \epsilon\right]$$

The first addend goes to zero by lemma A.3, the second addend equals zero, and the probability in the last addend goes to 1. Thus for $T$ large enough $\frac{1}{T} I(A^T \to B^T) \geq \underline{I} - 2\epsilon$. Since $\epsilon$ is arbitrary we see that $\underline{I}(A \to B) \leq \liminf_{T \to \infty} \frac{1}{T} I(A^T \to B^T)$.

Now we treat the last inequality. For any $\epsilon > 0$ we have

$$\frac{1}{T} I(A^T \to B^T) \leq E\left[\frac{1}{T} \log \frac{\vec{p}(B^T \mid A^T)}{P(B^T)} 1_{\left\{\frac{1}{T} \log \frac{\vec{p}(B^T \mid A^T)}{P(B^T)} \geq L\right\}}\right]$$

$$+ LP\left[L \geq \frac{1}{T} \log \frac{\vec{p}(B^T \mid A^T)}{P(B^T)} \geq \bar{I}(A \to B) + \epsilon\right]$$

$$+ (\bar{I}(A \to B) + \epsilon)P\left[\frac{1}{T} \log \frac{\vec{p}(B^T \mid A^T)}{P(B^T)} \leq \bar{I}(A \to B) + \epsilon\right]$$

The first addend goes to zero by lemma A.4, the second addend goes to zero by definition of $\bar{I}$, and the probability in the last addend goes to 1. Thus for $T$ large enough $\frac{1}{T} I(A^T \to B^T) \leq \bar{I} + 2\epsilon$. Since $\epsilon$ is arbitrary we see that $\limsup_{T \to \infty} \frac{1}{T} I(A^T \to B^T) \leq \underline{I}(A \to B)$. □

### A.3 Lemma 7.7

We repeat the statement of lemma 7.7 for convenience.
**Lemma 7.7** For $|\mathcal{B}|$ finite we have

(1) The cost is bounded and continuous. Specifically, $0 \leq \bar{c}(u) \leq \log|\mathcal{B}|$, $\forall u \in \mathcal{U}$.

(2) The control constraint function $\mathcal{U}(\gamma)$ is a continuous set-valued map between $\mathcal{P}(\mathcal{P}(\mathcal{S}))$ and $\mathcal{U}$.

(3) The dynamics $r(d\gamma_{t+1} \mid \gamma_t, u_t)$ are continuous.

**Proof:** To prove part (1) recall $\bar{c}(u) = \int r(db \mid \pi, a) u(d\pi, da) \log \frac{r(b \mid \pi, a)}{\int r(b \mid \tilde{\pi}, \tilde{a}) u(d\tilde{\pi}, d\tilde{a})}$. This $\bar{c}(u)$ corresponds to a mutual information with input distribution $u(d\pi, da)$ and an output $B$ in a finite alphabet $\mathcal{B}$. Hence $\bar{c}(u) \leq \log|\mathcal{B}|$ $\forall u \in \mathcal{U}$. The cost is clearly continuous in $u \in \mathcal{U}$.

To prove part (2) recall $\mathcal{U}(\gamma) = \{u(d\pi, da) : u(d\pi, da) \in \mathcal{U}, u(d\pi) = \gamma(d\pi)\}$. The set $U(\gamma)$ is compact for each $\gamma \in \mathcal{P}(\mathcal{P}(\mathcal{S}))$. For any set $H \subset \mathcal{U}$ denote $\mathcal{U}^{-1}(H) = \{\gamma : U(\gamma) \cap H \neq \emptyset\}$. The set-valued map $\mathcal{U}(\gamma)$ is *continuous* if it is both

(1) Upper semicontinuous (usc): $\mathcal{U}^{-1}(F)$ is closed in $\mathcal{P}(\mathcal{P}(\mathcal{S}))$ for every closed set $F \subset \mathcal{U}$.

(2) Lower semicontinuous (usc): $\mathcal{U}^{-1}(G)$ is open in $\mathcal{P}(\mathcal{P}(\mathcal{S}))$ for every open set $G \subset \mathcal{U}$.

The control constraint $U(\gamma)$ is clearly both usc and lsc and hence is continuous.

To prove part (3) recall equation (33):

$$r(d\gamma_{t+1} \mid \gamma_t, u_t) = \int_{\mathcal{U}} \int_{\mathcal{A}} \int_{\mathcal{B}} \delta_{\{\Phi_\Gamma(u_t, b_t)\}}(d\gamma_{t+1}) \, r(db_t \mid \pi_t, a_t) \, u_t(d\pi_t, da_t).$$

Since this stochastic kernel does not depend on $\gamma_t$ we only need to show that it is continuous in $u_t$. Specifically, let $v$ be any continuous bounded function on $\mathcal{P}(\mathcal{P}(\mathcal{S}))$. We need to show

$$\int \int \int v(\Phi_\Gamma(u,b)) \, r(db \mid \pi, a) \, u(d\pi, da) \tag{A1}$$

is a continuous function of $u_t$.

By equation (25) we know for all Borel measurable $\Omega \subset \mathcal{P}(\mathcal{S})$:

$$\gamma[u,b](\Omega) = \int \int \{\Phi_\Pi(\pi, a, b) \in \Omega\} r(d\pi, da \mid u, b). \tag{A2}$$

By lemma 7.1 we know for any Borel measurable $\Theta \subset \mathcal{P}(\mathcal{S}), a, b$, and $u$ we have

$$r(\Theta, a \mid u, b) = \frac{\int_\Theta r(b \mid \tilde{\pi}, a) \, u(d\tilde{\pi}, a)}{\int \int r(b \mid \tilde{\pi}, \tilde{a}) \, u(d\tilde{\pi}, d\tilde{a})} \tag{A3}$$

when the denominator does not equal zero. Because $\mathcal{B}$ is finite and by repeated use of Theorem A.4 we see that (A3) is continuous in $u, b$ for all $\Theta$. This implies (A2) is continuous in $u, b$ for all $\Omega$. Thus implying (A1) is continuous in $u$. $\square$